\documentclass[letterpaper,11pt]{article}
\pdfoutput=1 
\usepackage{jheppub} 
\usepackage[T1]{fontenc} 
\usepackage[dvipsnames]{xcolor}

\DeclareMathOperator{\Tr}{Tr}

\title{
QCD Vacuum in an Inhomogeneous Magnetic Field
}

\author[a,b]{Prabal Adhikari}
\author[c,d,e]{and Brian C. Tiburzi}

\affiliation[a]{Physics Department, Faculty of Natural Sciences and Mathematics, St.~Olaf College,\\
Northfield, Minnesota 55057, USA}
\affiliation[b]{Kavli Institute for Theoretical Physics, University of California,\\ 
Santa Barbara, California 93106, USA}
\affiliation[c]{Department of Physics, The City College of New York, \\ New York, New York 10031, USA}
\affiliation[d]{Graduate School and University Center, The City University of New York,\\
New York, New York 10016, USA}
\affiliation[e]{Institute for Advanced Simulation (IAS-4), Forschungszentrum J\"ulich,\\
52428 J\"ulich, Germany}

\emailAdd{prabal.adhikari.physics@proton.me}
\emailAdd{btiburzi@ccny.cuny.edu}

\abstract{The effect of an inhomogeneous magnetic field on the QCD vacuum is addressed using 
the framework of chiral perturbation theory. 
The magnetic field is chosen to be localized along one spatial direction, 
with a profile for which the underlying quantum mechanical problem is exactly solvable. 
Particular attention is paid to regularization and renormalization using dimensional regularization.
While the non-vanishing gradient of the magnetic field requires additional operators in chiral perturbation theory, 
their effect occurs at next-to-next-to-leading order in the chiral expansion. 
Consequently, 
the magnetic field dependence of equilibrium vacuum observables can be determined at next-to-leading order without undetermined parameters. 
We compute the zero-temperature free energy and chiral condensate for the inhomogeneous background, 
both as integrated quantities as well as spatially resolved local observables. 
Comparison with locally constant approximations enables a direct probe of the spatial response and nonlocal structure of the magnetized QCD vacuum.
We additionally derive the induced vacuum current associated with the inhomogeneity of the magnetic field.
}

\begin{document} 
\maketitle
\flushbottom

\def\a{{\alpha}}
\def\be{{\beta}}
\def\d{{\delta}}
\def\D{{\Delta}}
\def\P{{\Pi}}
\def\p{{\pi}}
\def\e{{\varepsilon}}
\def\ep{{\epsilon}}
\def\g{{\gamma}}
\def\G{{\Gamma}}
\def\k{{\kappa}}
\def\l{{\lambda}}
\def\L{{\Lambda}}
\def\m{{\mu}}
\def\n{{\nu}}
\def\o{{\omega}}
\def\O{{\Omega}}
\def\S{{\Sigma}}
\def\s{{\sigma}}
\def\t{{\tau}}
\def\x{{\xi}}
\def\X{{\Xi}}
\def\z{{\zeta}}
\def\ol#1{{\overline{#1}}}
\def\c#1{{\mathcal{#1}}}

\section{Introduction}
\label{sec:intro}

The study of quantum field theories in external fields was pioneered long ago by Schwinger%
~\cite{Schwinger:1951nm}, 
who established many of the conceptual and technical foundations for understanding vacuum polarization 
and the response of quantum fluctuations to classical background fields.
These methods, 
together with numerous subsequent developments, 
have become standard tools for investigating non-perturbative aspects of quantum field theory,
and are reviewed in
Refs.~\cite{Dittrich:2000zu,Dunne:2004nc,Miransky:2015ava}. 
External electromagnetic fields provide a particularly useful probe, 
because they couple directly to charged degrees of freedom 
and thereby reveal the structure of the vacuum through its induced response.

In the context of quantum chromodynamics (QCD), 
there has been considerable interest in understanding the response of 
strong interactions to external electromagnetic fields. 
Such studies are motivated by the QCD phase diagram%
~\cite{Kharzeev:2013jha,Andersen:2014xxa,Adhikari:2024bfa}, 
and by the intense electromagnetic fields generated in heavy-ion collisions, 
which may reach hadronic scales during the earliest stages of the collision%
~\cite{Skokov:2009qp}. 
External magnetic fields have also been considered in a variety of astrophysical settings, 
including magnetars, where field strengths far exceed those achievable in terrestrial laboratories%
~\cite{Harding:2006qn}. 
Consequently, 
electromagnetic fields provide a valuable means of probing both the structure of the QCD vacuum and the properties of strongly interacting matter under extreme conditions.
A wealth of first-principles calculations of QCD in external electromagnetic fields has been performed using lattice gauge theory. These studies have provided quantitative information on magnetic catalysis, magnetic susceptibilities, equation-of-state properties, and the behavior of hadronic observables in external fields; 
see 
Ref.~\cite{Endrodi:2024cqn}
for a comprehensive overview. 
Complementary insight has been obtained from chiral perturbation theory%
~\cite{Gasser:1983yg,Gasser:1984gg}, 
which provides a systematic framework for describing the low-energy dynamics of the QCD vacuum 
and low-lying hadronic degrees of freedom, 
including the response of QCD to external electromagnetic fields.

Most theoretical investigations have focused on uniform electromagnetic fields. 
In this case, 
translational invariance is preserved, 
considerably simplifying both analytical and numerical calculations. 
Physical electromagnetic fields, however, are generally neither uniform nor static. 
The magnetic fields produced in heavy-ion collisions vary rapidly in both space and time, 
while astrophysical magnetic fields often exhibit significant spatial structure. 
Such inhomogeneities introduce new length scales 
and permit nonlocal responses that are absent in uniform backgrounds. 
Understanding how the QCD vacuum responds to spatially varying electromagnetic fields 
is therefore of interest in its own right
and provides a natural extension of the extensively studied uniform-field problem.
Analytic insight can nevertheless be obtained from external-field profiles that are solvable in closed form. 
Such investigations were undertaken in a series of works developing exact methods for evaluating quantum effective actions in solvable electromagnetic backgrounds%
~\cite{Cangemi:1995ee,Dunne:1995cj,Dunne:1997kw,Dunne:1998ni,Dunne:1999uy}. 
Connection to QCD was subsequently sought through a model study
employing the Nambu--Jona-Lasinio model in an inhomogeneous magnetic field%
~\cite{Cao:2017gqs}. 
While offering qualitative insight, 
the analysis is necessarily model dependent and relies on further approximations.
More recently, 
first-principles calculations directly from QCD have now been carried out using 
lattice gauge theory techniques%
~\cite{Brandt:2023dir}, 
including an investigation of the induced electric current 
\cite{Brandt:2024blb}. 
The goal of the present work is to provide a closed-form treatment of this problem within chiral perturbation theory, 
thereby obtaining the corresponding low-energy effective field theory description 
of the QCD vacuum response to this solvable class of inhomogeneous magnetic fields.

Our presentation is organized as follows. 
First in Section~\ref{sec:inhomo}, 
we review chiral perturbation theory in an external magnetic field, 
focusing on the computation of  
the effective action in the presence of inhomogeneous magnetic fields that vary along one spatial direction. 
Pertinent terms from the chiral Lagrangian are identified, 
with consideration given to regularization and renormalization. 
The mathematically degenerate case of a uniform magnetic field 
is revisited to demonstrate that familiar results for
the free energy, chiral condensate, and magnetization are recovered. 
In Section~\ref{sec:bulk}, 
integrated vacuum observables
are obtained for a particular solvable profile of a 
non-uniform magnetic field. 
Despite the field inhomogeneity, 
no additional terms from the chiral Lagrangian enter at next-to-leading order. 
We show that the counter-term determined from the uniform field problem accordingly ensures 
the proper renormalization of the non-uniform field problem. 
The integrated chiral condensate and integrated renormalized magnetization are explored 
as a function of the width of the magnetic field. 
Spatially resolved observables are considered in 
Section~\ref{sec:local}, 
through the local free-energy density.
The spatial profile of the chiral condensate is investigated
and used in comparison with locally constant approximations to probe the
nonlocal response of the QCD vacuum to spatially varying magnetic fields. 
Additionally, 
we determine the induced vacuum current generated by the magnetic-field gradient, 
and examine its relation to the local magnetic response. 
A few concluding remarks in Section~\ref{sec:conc} end our work.

\section{Chiral Perturbation Theory in an Inhomogeneous Magnetic Field}
\label{sec:inhomo}

\subsection{Chiral Lagrangian and Effective Action}

Low-energy QCD can be described in a model-independent fashion using 
chiral perturbation theory. 
This effective theory is organized in powers of the pion mass and momentum. 
Additionally, 
external sources can be included, 
of which we focus on static magnetic fields. 
The inhomogeneous magnetic fields considered throughout have the general form 
\begin{equation}
\vec{B} = B(x) \, \hat{z}
\label{eq:Bgen}
,\end{equation} 
where we have chosen to align the field with the 
$z$ axis, 
and have taken the localization of the magnetic field to be along the 
$x$ axis.  
While a particular profile of the magnetic field
$B(x)$
is chosen below in Eq.~\eqref{eq:Bsech}
based on solvability, 
we first highlight features of this more general problem, 
leaving the profile function temporarily unspecified. 
A convenient gauge in which to implement such a magnetic field is
$\vec{A} = A(x) \, \hat{y}$, 
where 
$A(x) = \int B(x) \, dx$
and the integration constant is pure gauge. 
In this asymmetric gauge, 
a charged particle has 
$d{-}1$
good components of momentum, 
where 
$d$
is the number of spacetime dimensions. 
As a matter of convenience, 
we work in 
$d$-dimensional Euclidean space throughout.  
In employing dimensional regularization, 
we shall take 
$d = 4 {-} 2 \e$
and analytically continue to 
$\e \ll 1$, 
with the fourth dimension corresponding to Euclidean time.

Central to the computations that follow is the effective action, 
which is the logarithm of the QCD partition function. 
In the effective field theory, 
the partition function is determined order-by-order in the chiral expansion. 
It can be written in terms of a functional integral over the pion fields
\begin{equation}
\c Z
[A]
=
\int \c D \p^+ \c D \p^- 
e^{ - \int d^d x \, \c L_E}
,\end{equation}
where 
$\c L_E$
is the Euclidean Lagrangian density, 
and we focus only on the charged-pion contribution. 
The neutral pion does not contribute to the magnetic free energy at 
next-to-leading order due to charge neutrality. 
The magnetic free-energy density 
$\c F(x)$
can be defined%
\footnote{
Unlike homogeneous systems, 
there is generally an ambiguity in identifying a local free-energy density from the effective action.
For a spatial profile depending on a single coordinate, 
different local representations of the free-energy density are related by a total derivative 
$\c F(x) \to \c F (x) + \frac{d}{dx} \phi(x)$, 
where the boundary condition 
$\phi ( + \infty ) = \phi (- \infty )$
ensures that the effective action remains unchanged.
In what follows, 
we adopt the local representation naturally obtained from the exact resolvent form of the effective action. 
} 
from the spatial integrand of the effective action 
\begin{equation}
\Gamma
[A]
- 
\Gamma
[0]
= 
- \log \frac{\c Z[A]}{\c Z[0]}
= 
\int \c F(x) \, d^dx 
\label{eq:F}
.\end{equation}
As the effective action difference
vanishes in zero magnetic field, 
the free-energy density 
$\c F(x)$
represents contributions that depend on the magnetic field.  
Note that with the magnetic free-energy density defined as in 
Eq.~\eqref{eq:F}, 
any next-to-leading order contributions from the neutral-pion fields necessarily cancel in the difference.

At leading order in the chiral expansion%
~\cite{Gasser:1983yg}, 
the Euclidean Lagrangian density including the Maxwell term for the external field reads
\begin{equation}
\c L_E^{(2)}
=
\frac{B^2(x)}{2}  
+
\frac{F^2}{4}
\Big\langle
D_\m U^\dagger  D_\m U
\Big\rangle
+
\frac{\S_0}{4}
\Big\langle
s(x)
( U^\dagger {+} U)
\Big\rangle
,\end{equation}
where angled brackets denote traces over flavor. 
The parameter
$F$
is the chiral-limit value of the pion decay constant. 
We include a scalar source 
$s(x)$, 
and the low-energy constant 
$\S_0 {<} \, 0$
accompanying this term is the chiral-limit value of the chiral condensate. 
The pseudo-Goldstone pion fields
$\vec{\p}$
are embedded in the coset field 
$U$
in the form
$U = \exp( i \, \vec{\p} \cdot \vec{\t} / F)$, 
where 
$\vec{\tau}$
are isospin matrices. 
At leading order, 
the pions couple to the external magnetic field through the gauge covariant derivative, 
whose action is specified by
\begin{equation}
D_\m U = \partial_\m U + i A_\m \left[ \c Q, U \right]
,\end{equation}
in the absence of other vector source fields. 
The quark electric charge matrix 
$\c Q$
contains both isoscalar and isovector terms
\begin{equation}
\c Q = e \left( \frac{1}{6} + \frac{1}{2} \tau^3 \right)
,\end{equation} 
where
$e > 0$
is the electric charge of the proton. 
Expanding about the vacuum expectation value of the coset field
$U = 1 + \cdots$
and replacing the scalar source with the quark mass
$s(x) \to m_q$, 
the action density becomes constant
$\c L_E^{(2)} = m_q \, \S_0  + \cdots$. 
Due to its magnetic field independence,  
the leading-order term does not contribute to the magnetic free energy.

Modifications to the vanishing magnetic field limit are obtained by retaining the leading pion fluctuations about the 
vacuum value of 
$U$. 
Additionally, 
there are further operators required from the next-to-leading order Lagrangian. 
At next-to-leading order in the chiral expansion,
there are three terms relevant for the magnetic free energy. 
These appear in the Euclidean Lagrangian density
\begin{equation}
\c L_E^{(4)} 
\supset
- 
l_5 \big \langle U^\dagger \hat{R}_{\m \n} U \hat{L}_{\m \n} \big \rangle
+ 
\tfrac{1}{2} (l_5 + 4 h_2) \big \langle \hat{R}_{\m \n} \hat{R}_{\m \n} + \hat{L}_{\m \n} \hat{L}_{\m \n} \big \rangle
-
\tfrac{1}{4} h_4 \big \langle R_{\m \n} + L_{\m \n} \big \rangle \big \langle R_{\m \n} + L_{\m \n} \big \rangle
\label{eq:L4}
,\end{equation} 
where 
$R_{\m \n}$
and
$L_{\m \n}$
are right- and left-handed field-strength tensors. 
Those with hats are defined to be traceless 
$\hat{O} = O - \frac{1}{2} \langle O \rangle$, 
which then correspond to the isovector contributions.  
Using a different representation of the coset field, 
the operators with coefficients 
$l_5$
and
$h_2$
were given in Ref.~\cite{Gasser:1983yg}, 
while that with coefficient 
$h_4$
was not considered due to the exclusion of isoscalar gauge fields.
This contact operator appears, for example, 
in Ref.~\cite{Kaiser:2000ck}, 
or can be generated from matching the three-flavor chiral Lagrangian%
~\cite{Gasser:1984gg}
to the two-flavor one. 
In the present case of a magnetic field, 
we set
$R_{\m \n} = L_{\m \n} = - \c Q F_{\m \n}$, 
where 
$F_{\m \n} = \partial_\m A_\n - \partial_\n A_\m$
is the electromagnetic field-strength tensor.

In the external magnetic field
Eq.~\eqref{eq:Bgen}, 
the Euclidean Lagrangian density to next-to-leading order accuracy is given by 
\begin{equation}
\c L_E
=
Z  \, \frac{B^2(x)}{2}  
-
\p^+ D_\m D_\m \p^- 
- \frac{s(x) \, \S_0}{F^2}
\, \p^+ \p^- 
+
\cdots
,\end{equation}
where the omitted terms do not depend on the magnetic field, or are 
suppressed by powers of 
$F^{-2}$. 
The field-strength renormalization is given by 
\begin{equation}
Z 
= 
1 {+} e^2  h
,\end{equation}
and depends on the parameter 
$h$, 
which is a linear combination of the low-energy constants from the next-to-leading order Lagrangian density
Eq.~\eqref{eq:L4}, 
specifically
\begin{equation}
h = 8 h_2 - \tfrac{4}{9} h_4
,\end{equation}
for which the contribution from 
$l_5$
cancels. 
The gauge covariant derivative of the charged pion field is specified in the usual way 
$D_\m \p^- = (\partial_\m - i e A_\m) \p^-$. 
The charged pions are canonically normalized, 
and their mass-squared satisfies the Gell-Mann--Oakes--Renner relation, 
namely
$m_q \, \S_0 = -  F^2 m^2$. 
For ease of notation, 
we will write 
$m^2(x) = - s(x) \S_0 / F^2$, 
which becomes the pion mass-squared when 
$s(x) \to m_q$. 
The functional integral over the charged scalar fields 
formally produces the partition function 
\begin{equation}
\c Z [A]
=
\z[A] \, e^{ - \Tr \log \big[ - D_\m D_\m + m^2(x) \big]}
,\end{equation}
where 
$\z[A]$
contains the contribution from the energy density of the magnetic field. 
In vanishing magnetic field, 
$\z[0]$
is an infinite constant that appears as a multiplicative factor
\begin{equation}
\z [A]
= 
\z [0] \,
\exp \left[ - \tfrac{1}{2} Z \int  B^2(x) \, d^dx \right]
.\end{equation}
Consequently, 
the effective action difference in
Eq.~\eqref{eq:F}
is given by the expression
\begin{equation}
\G[A] - \G[0]
=
\tfrac{1}{2} Z \int B^2(x) \, d^dx 
+
\Tr \log \left[ \frac{- D_\m D_\m + m^2(x)}{- \partial_\m \partial_\m + m^2(x)} \right]
\label{eq:FlogD}
,\end{equation}
which is the starting point for the computations that follow.

\subsubsection{Free-Energy Density}

To compute the free-energy density, 
we evaluate the effective action difference 
with the scalar source replaced by the quark mass. 
For an operator 
$\mathfrak{D}$, 
the logarithm of its functional determinant is defined presently using dimensional regularization
\begin{equation}
\log \det \mathfrak{D}
=
\Tr \log \mathfrak{D}
=
\int d^d x  \, \langle x_\m | \log \mathfrak{D} \, | x_\m \rangle
.\end{equation}
With the gauge choice used to implement the magnetic field 
Eq.~\eqref{eq:Bgen}, 
the differential operator 
$\mathfrak{D}$
in a magnetic field is of the general form
\begin{equation}
\mathfrak{D}
=
- D_\m D_\m + m^2
=
- 
\frac{\partial^2}{\partial x^2}
- 
\left[ \frac{\partial}{\partial y} - i e A(x) \right]^2
- 
\left( \nabla^2 \right)_{d-2}
+
m^2
,\end{equation}
where 
$\left( \nabla^2 \right)_{d-2}$
is the 
$(d{-}2)$-dimensional Laplacian. 
As there are 
$d{-}1$
good components of momentum, 
the operator
$\mathfrak{D}$
can be partially diagonalized in momentum space. 
To this end, 
we write the 
$(d{-}1)$-dimensional momentum-space matrix elements as
\begin{equation}
\mathfrak{D} (p_2, E)
= 
- \frac{d^2}{dx^2} + V(p_2,x) + E 
,\end{equation}
where the one-dimensional potential has the form 
\begin{equation}
V(p_2,x) = \Big[ p_2 - e A(x) \Big]^2 + m^2
,\end{equation} 
and depends on the $y$-component of momentum
$p_2$. 
The parameter 
$E$
plays the mathematical role of the (negative) energy in the effective one-dimensional Schr\"odinger equation, 
and results from the action of the 
$(d{-}2)$-dimensional Laplacian, 
so that
$E = (p_3)^2 + (p_4)^2 + \cdots$.  
This leads to mixed spatial and momentum integration in the trace
\begin{equation}
\Tr \log \mathfrak{D}
=
\frac{2}{(4\p)^{\frac{d}{2}} \G\left(\frac{d-2}{2}\right)}
\int d^dx
\int_{-\infty}^{+\infty} dp_2
\int_0^\infty d E \, E^{\frac{d-4}{2}} \,
\big\langle x \big| \log \mathfrak{D} (p_2, E) \, \big| x \big\rangle
\label{eq:logD}
,\end{equation}
having appealed to 
$SO(d{-}2)$
rotational invariance for the 
$d{-}2$
integrals over components of the momentum
$p_3$, 
$p_4$,
$\cdots$.

To avoid computation of the logarithm of 
$\mathfrak{D}$, 
we trivially introduce a derivative by writing the factor 
$E^{\frac{d-4}{2}}$ 
as
$\left( \frac{d-2}{2} \right)^{-1} \frac{\partial}{\partial E} \, E^{\frac{d-2}{2}}$, 
which enables an integration by parts to be performed. 
The subsequently produced end-point contributions to the effective action difference in 
Eq.~\eqref{eq:FlogD}
are of the form
\begin{equation}
\propto \, 
E^{\frac{d-2}{2}} 
\Big[ 
\log \mathfrak{D} (p_2 , E)
-
\log \mathfrak{D}_0 (p_2 ,E)
\Big]
\Bigg|_{E = 0}^{E = \infty}
=
0
,\end{equation} 
where 
$\frak{D}_0 (p_2, E) = - \frac{d^2}{dx^2} + p_2^2 + m^2 + E$
is the corresponding operator in zero magnetic field. 
Each term vanishes at the lower endpoint provided 
$d > 2$.
At the upper endpoint, 
each logarithm has the large 
$E$ 
behavior 
$\sim \log E$
independent of the magnetic field
and exactly cancels in the difference of logarithms. 
The next-order term from each logarithm for 
$E \gg 1$
is proportional to 
$E^{-1}$. 
This leads to an overall
$E^{\frac{d-4}{2}}$
when combined with the multiplicative pre-factor,
and 
consequently vanishes provided that
$d < 4$.
After integration by parts, 
the derivative of the logarithm produces the Green's function of the effective one-dimensional problem
\begin{equation}
\frac{\partial}{\partial E}
\big\langle x' \big| \log \mathfrak{D} (p_2, E) \, \big| x \big\rangle
=
\big\langle x' \big| \frac{1}{- \frac{d^2}{dx^2} + V(p_2,x) + E} \big| x  \big\rangle
=
G(x',x|p_2, E)
\label{eq:1dSch}
.\end{equation}
For the trace, 
only the coincident Green's function
is required,
for which 
$x' = x$. 
The representation of the local magnetic free-energy density defined in 
Eq.~\eqref{eq:F}
can thus be obtained from the formula
\begin{equation}
\c F(x)
=
Z
\, \frac{B^2(x)}{2}  
- 
\frac{2}{(4\p)^{\frac{d}{2}} \G\left(\frac{d}{2}\right)}
\int_{-\infty}^{+\infty} d p_2 \int_0^\infty d E \, E^{\frac{d-2}{2}}
\Big[
G(x,x|p_2, E)
{-}
G_0(0,0|p_2, E)
\Big]
\label{eq:flocal}
.\end{equation}
We use the notation 
$G_0$
for the corresponding 
Green's function in zero magnetic field. 
This zero-field Green's function is translationally invariant. 
The dimensionally regulated integral must be evaluated with care to isolate divergent contributions, 
and the counter-term 
$h$
from the chiral Lagrangian must accordingly be determined.

\subsubsection{Chiral Condensate and Vacuum Current}

In addition to the free energy, 
the chiral condensate and vacuum current are investigated. 
The former arises from the functional derivative of the effective action with respect to the scalar source
\begin{equation}
\S(x)
= \frac{\d \, \G[A]}{\d s(x)}
\Bigg|_{s(x) = m_q}
.\end{equation}
Unlike the free energy, 
there is no ambiguity in defining this local response;
however, 
it is renormalization scale and scheme dependent. 
A renormalization group invariant quantity is the relative difference in the chiral condensate compared 
to its zero-field value
$\S$. 
The condensate difference can be computed from the functional derivative of the effective action difference in 
Eq.~\eqref{eq:FlogD}, 
namely
\begin{equation}
\S(x) - \S
=
-\frac{\S_0}{F^2}
\Big[ 
\frak{D}^{-1}(x,x) 
- 
\frak{D}^{-1}_0(0,0)
\Big]
.\end{equation}
In an inhomogeneous magnetic field, 
the coincident propagator is only a function of 
$x$ 
due to 
$(d{-}1)$-dimensional translational invariance, 
while the zero-field propagator is fully translationally invariant. 
Using the mixed space-momentum Green's function, 
the relative difference in the chiral condensate is given to next-to-leading order accuracy by
\begin{equation}
\frac{\S(x) {-} \S}{\,\, \S}
=
- \frac{2}{(4\p)^{\frac{d}{2}} \G \left( \frac{d-2}{2} \right) F^2}
\int_{-\infty}^{+\infty} dp_2
\int_0^\infty d E
\, E^{\frac{d-4}{2}}
\Big[
G(x,x|p_2, E)
-
G_0(0,0|p_2, E)
\Big]
\label{eq:condD}
.\end{equation}
Unlike the free-energy density, 
the condensate difference is finite in the ultraviolet.

The local response of the QCD vacuum can also be studied using the induced vacuum current. 
It can be obtained from the functional derivative of the effective action with respect to the vector source field 
\begin{equation}
J_\m(x)
=
-
\frac{\d \, \G[\c A]}{\d \c A_\m (x)}
\Bigg|_{\c A_\m = A_\m}
,\end{equation}
where the subsequent evaluation is carried out in the inhomogeneous magnetic background
Eq.~\eqref{eq:Bgen}. 
The inhomogeneous magnetic field requires a source current, 
while the magnetization of the vacuum
$\vec{\c M}$
subsequently induces a vacuum current. 
These contributions are contained in the spatial components of the current vector  
\begin{equation}
\vec{J}
= 
\vec{\nabla}
{\times}
\left( 
\vec{B}
{-} 
\vec{\c M}
\right)
.\end{equation}
Similar to the chiral condensate, 
there is no ambiguity in the local vacuum current. 
Using the effective action 
Eq.~\eqref{eq:FlogD}, 
we obtain the general expression for the vacuum current in terms of the propagator
\begin{equation}
\vec{J}(x)
=
Z \, \vec{\nabla} {\times} \vec{B}(x)
+
i e 
\lim_{x'_\m \to \, x_\m}
\left(
\vec{D}' {-} \vec{D}^*
\right)
\frak{D}^{-1}(x',x)
\label{eq:currentgen}
.\end{equation}
Due to the form of the magnetic field in 
Eq.~\eqref{eq:Bgen}, 
only the 
$y$-component of the current is non-vanishing. 
The point splitting of the current is required to guarantee the vanishing of the current's 
$x$-component. 
While the vacuum current is divergent in the ultraviolet, 
the divergence is canceled by the counter-term in the source current. 
Accordingly, 
the sum of these two contributions renders 
$\vec{J}$ 
finite.

Before considering the case of an inhomogeneous magnetic field, 
we show that familiar results for the chiral condensate and magnetization are recovered for a uniform 
magnetic field. 
Thereby, 
we also determine the counter-term.

\subsection{Degenerate Example: Uniform Magnetic Field}

While the development above is sufficiently general to handle a magnetic field depending on one coordinate, 
it is useful to verify that results expected for a uniform magnetic field are correctly recovered. 
To this end, 
we consider this mathematically degenerate case,
for which
$B(x) = B$
is uniform and the gauge potential is a linear function
$A(x) = B x$. 
The corresponding one-dimensional potential is that of a shifted simple harmonic oscillator
\begin{equation}
V(p_2,x)
= 
(p_2 - e B x)^2 + m^2
\label{eq:VSHO}
.\end{equation}
To compute the magnetic free-energy density, 
we use Schwinger's proper-time integral, 
which is closely related to the spectral representation of the Green's function.

\subsubsection{Free Energy and Renormalization}

To compute the free-energy density 
Eq.~\eqref{eq:flocal}, 
we need the trace of the one-dimensional Green's function. 
With the proper-time integral representation, 
the coincident Green's function is
\begin{align}
G(x,x | p_2, E)
&=
\int_0^\infty ds \, e^{ - s \, E}
\, \langle x |
e^{ - s \left[ - \frac{d^2}{dx^2} + V(p_2,x) \right]}
| x \rangle
\notag \\
&=
\int_0^\infty ds \, e^{ - s \left( E + m^2 \right)}
\sqrt{\frac{eB}{2 \p \sinh 2 e B s}}
\, e^{ - e B \tanh e B s \, \left( x - \frac{p_2}{e B}\right)^2 }
\label{eq:Schwinger}
,\end{align}
where the well-known quantum mechanical propagator for the harmonic oscillator has been utilized in the second line. 
In the expression for the magnetic free-energy density, 
the integral over 
$E$
is an instance of the gamma-function, 
while that over 
$p_2$
is Gaussian. 
In the latter integral, 
the $x$ dependence can be translated away via the shift of integration
$p_2 \to p_2 + e B x$. 
Resulting quantities are independent of 
$x$, 
as expected for a uniform field. 
For this reason, 
we denote the free-energy density as 
$\c F$
in the uniform field case, 
rather than 
$\c F(x)$.

Carrying out similar but simpler integrals for the zero-field Green's function 
$G_{0}(0,0|p_2, E)$, 
we arrive at the magnetic free-energy density in the form
\begin{equation}
\c F
=
Z \, 
\frac{B^2}{2} 
- 
\frac{1}{(4\p)^{\frac{d}{2}}}
\int_0^\infty \frac{ds}{s^{\frac{d+2}{2}}}
e^{- s m^2}
\left( 
\frac{e B s}{\sinh e B s} - 1
\right)
,\end{equation}
which accordingly vanishes for 
$B = 0$. 
Near 
$d {=} 4$
dimensions, 
the proper-time integral contains an ultraviolet divergence
$\sim \log s$,
for 
$s \ll 1$. 
The divergence can be isolated by adding and subtracting the first non-trivial term in the 
weak-field expansion of the integrand. 
This procedure results in the expression
\begin{equation}
\c F
=
\left\{
1 
+ 
e^2 
\left[
\frac{\G(\e)}{3 (4 \p)^2}
\left( \frac{4 \p}{m^2} \right)^{\e}
+
h
\right]
\right\}
\frac{B^2}{2}
+ 
\c F^r_B
\label{eq:Fnon}
,\end{equation}
for which the renormalized integral 
$\c F^r_B$
is now finite in 
$d {=} 4$ 
dimensions
and given by the formula
\begin{equation}
\c F^r_B
=
- m^4 \int_0^\infty \frac{ds}{(4 \p)^2 s^{3}} \, e^{- s}
\left( 
\frac{\eta s}{\sinh \eta s} - 1 + \frac{(\eta s)^2}{6}
\right)
\label{eq:FB}
,\end{equation}
where 
\begin{equation}
\eta = \frac{e B}{m^2}
,\end{equation}
is a parameter employed throughout.

With the divergence isolated near 
$d {=} 4$
dimensions, 
we can define a renormalized running coupling 
$h^r(\L)$
using the modified minimal subtraction 
($\overline{\text{MS}}$)
scheme%
\footnote{
The one-loop divergence is exactly canceled by the known counter-term 
$h_2$%
~\cite{Gasser:1983yg}, 
because the isoscalar component of the electromagnetic field does not contribute at this order. 
Consequently,
$h_4$
must be a finite counter-term at next-to-leading order. 
}
\begin{equation}
h
=
- \frac{1}{3(4\p)^2} 
\left( \frac{1}{\e} - \g_E + \log \frac{4\p}{\, \L^2} \right)
+
h^r(\L)
\label{eq:h}
.\end{equation}
Additionally, 
we can define a renormalization group invariant coupling 
$\ol h$
through the relation
\begin{equation}
\ol h 
= 
3 (4 \p)^2 h^r(\L) - \log \frac{m^2}{\L^2}
,\end{equation}
which is a more customary choice in two-flavor chiral perturbation theory. 
Rewriting the magnetic free-energy density 
Eq.~\eqref{eq:Fnon}
in terms of the renormalization group invariant coupling, 
we have
\begin{equation} 
\c F
= 
\frac{1}{2}
\left[
1 + \frac{e^2 \, \ol h}{3 (4 \p)^2}
\right]
B^2
+
\c F^r_B 
\label{eq:Fren}
.\end{equation}

\subsubsection{Chiral Condensate and Renormalized Magnetization}

In a uniform magnetic field, 
the chiral condensate is also uniform and we denote it by 
$\S_B$
rather than 
$\S(x)$. 
From Eq.~\eqref{eq:condD},%
\footnote{
The chiral condensate can also be determined from the quark-mass derivative of the free-energy density%
~\cite{Adhikari:2023fdl}, 
specifically 
$\frac{\S_B {-} \S}{\S}
=
- \frac{1}{F^2} \frac{\partial \c F}{\partial m^2}$.
Starting with 
Eq.~\eqref{eq:Fren}, 
one arrives at the same result provided the pion-mass dependence of the low-energy constant
$\ol h$
is accounted for, 
namely
$\frac{\partial}{\partial m^2} \ol h = - \frac{1}{m^2}$.
}
we must perform two integrals over the Green's function in Eq.~\eqref{eq:Schwinger}.  
Similar to above, 
the 
$E$ 
integral is an instance of the gamma-function
and the momentum integral is Gaussian. 
The relative difference in the chiral condensate is thus given by 
\begin{equation}
\frac{\S_B {-}  \S}{\S}
= 
-
\frac{m^2}{(4 \p F)^2}
\int_0^\infty \frac{ds}{s^2} \, e^{- s}
\left( \frac{\eta s}{\sinh \eta s} - 1 \right)
,\end{equation}
and appropriately reproduces the next-to-leading order chiral perturbation theory result in a uniform 
magnetic field%
~\cite{Cohen:2007bt}, 
which can be expressed as
\begin{equation}
\frac{\S_B {-}  \S}{\S}
=
- 
\frac{m^2}{(4 \p F)^2} \
\eta \,
\c I \left( \eta^{-1} \right)
\label{eq:chiralcond}
,\end{equation}
where
\begin{equation}
\c I(z)
=
2 \log \, \Gamma\left( \tfrac{1+z}{2} \right) 
+ 
z
\left( 1 - \log \tfrac{z}{2} \right)
- 
\log 2 \p
.\end{equation}

For uniform systems, 
the magnetization
$\vec{M}$
is also uniform; consequently,  
there is no vacuum current
$\vec{\nabla} \times \vec{H} = 0$. 
Instead, 
the magnetization is obtained from the derivative of the free-energy density with respect to the 
homogeneous magnetic field, 
specifically in the form 
\begin{equation}
\vec{H}
= 
\frac{1}{\m_0} \vec{B} - \vec{\c M}
=
\left( \frac{\partial \c F}{\partial \vec{B}} \right)_V
\label{eq:H}
.\end{equation}
Although we work in units where the bare magnetic permeability of the vacuum is
$(\m_0)_\text{bare} = 1$, 
the magnetic free-energy density in 
Eq.~\eqref{eq:Fren}
produces the low-energy QCD contribution to the permeability
\begin{equation}
\frac{\m_{0}}{(\m_0)_\text{bare}}
=
1 - \frac{\a_\text{fs}}{12 \p} \, \ol h + \cdots
\label{eq:mubare}
.\end{equation}
Despite the appearance of the fine-structure constant
$\a_\text{fs} = \frac{e^2}{4\p}$,  
this is not a QED effect. 
Instead, 
the effect owes to QCD vacuum fluctuations in a classical magnetic field. 
In principle, 
it can be extracted from lattice calculations, 
with the strong caveat that it must be QCD renormalization scale and scheme dependent.%
\footnote{
Despite the additional impossibility to measure this parameter experimentally, 
there is physics underlying its value, 
for example, 
$\ol h$
contains a chiral logarithm of relevance to the chiral condensate. 
The vacuum fluctuations in a magnetic field, 
moreover, 
are dominated by vector mesons, 
for which we estimate the value
$\ol h \approx 49$
using the results of resonance saturation calculations%
~\cite{Ecker:1988te}. 
The response of the QCD vacuum to a uniform magnetic field is diamagnetic at a scale 
$\Lambda \approx m_\rho$. 
}

For homogeneous systems, 
the effect can alternately be framed in terms of the magnetic susceptibility of the vacuum defined by
\begin{equation}
\chi_B
=
- \frac{\partial^2 \c F}{\partial (eB)^2} \Big|_{B=0} 
,\end{equation}
which is additively and not multiplicatively renormalized in QCD. 
The magnetic permeability is simply related to this parameter through 
$\m_0 = (1 {-} e^2 \chi_B)^{-1}$, 
but is additionally subject to multiplicative renormalization.
For a uniform magnetic field, 
the susceptibility depends only on the unknown coupling
\begin{equation}
\chi_{B}
= 
- \frac{\ol h}{3(4 \p)^2}
\label{eq:chiB}
.\end{equation}
In an inhomogeneous magnetic field, 
the susceptibility becomes a non-local response function 
$\chi_B(x,x')$, 
which we do not attempt to obtain in this work.
We will define a field-profile susceptibility below in 
Section~\ref{s:IFER}, 
where we consider the integrated free-energy density.

In a uniform magnetic field, 
the QCD renormalization of the permeability can be dispelled with by enforcing a simultaneous rescaling 
of the charge and classical magnetic field%
~\cite{Schwinger:1951nm}. 
With 
\begin{equation}
B \to \left(1 + \frac{\a_\text{fs}}{24 \p} \ol h \right)^{-\frac{1}{2}} B
\quad \text{and} \quad
e \to \left(1 + \frac{\a_\text{fs}}{24 \p} \ol h \right)^{\frac{1}{2}} e
\label{eq:rescale}
,\end{equation} 
one obtains the Maxwell contribution to the free-energy density
$\frac{1}{2} B^2$
with the product 
$e B$
unchanged. 
When computing the free-energy density in inhomogeneous magnetic fields,
we will not employ such rescaling of the charge and magnetic field. 
Instead, 
we will work with the modified Maxwell contribution and exhibit the cancelation of ultraviolet divergences 
in the free-energy density and vacuum current using the known counter-term
$h$. 
As we will see, 
there are additional finite contributions proportional to the square of the magnetic field strength in the case of the free-energy density, 
and linear in the magnetic field strength in the case of the vacuum current.

Regardless of how one handles the QCD renormalization of the Maxwell energy density, 
the renormalized magnetization in the case of a uniform field is obtained from the formula
\begin{equation}
\c M^r
=
- \left( \frac{\partial \c F^r_B}{\partial B} \right)_V
\label{eq:M}
,\end{equation}
as a consequence of 
Eqs.~\eqref{eq:Fren} 
and 
\eqref{eq:H}. 
As such, 
the renormalized magnetization in chiral perturbation theory stems only from currents generated from 
charged-pion vacuum fluctuations. 
This formula applied to 
Eq.~\eqref{eq:FB}
correctly produces the renormalized magnetization of low-energy QCD in a uniform magnetic field.

\section{Inhomogeneous Magnetic Field: Integrated Response}
\label{sec:bulk}

\subsection{An Inhomogeneous Magnetic Field}

An exactly solvable system with an inhomogeneous magnetic field is specified by the spatial profile%
~\cite{Cangemi:1995ee}
\begin{equation}
B(x) 
= 
B \, \text{sech}^2 \, \frac{x}{\l}
\label{eq:Bsech}
,\end{equation}
which can be implemented using the gauge function 
$A(x) = \l B \, \tanh \frac{x}{\l}$. 
In addition to the overall strength of the magnetic field  
$B$, 
there is the parameter 
$\l$
that determines the width of the field.  
Following the general discussion above, 
the magnetic free-energy density requires the Green's function 
Eq.~\eqref{eq:1dSch}
for the resulting effective one-dimensional Schr\"odinger equation. 
The corresponding potential can be written in the form  
\begin{equation}
V(p_2,x) 
= 
- 
\frac{\g^2 - \frac{1}{4}}{\l^2}
\left( 1- \tanh^2 \frac{x}{\l} \right)
+ 
\frac{\k_-^2}{2}  \left( 1 + \tanh \frac{x}{\l} \right)
+
\frac{\k_+^2}{2}  \left( 1- \tanh \frac{x}{\l} \right)
.\end{equation}
The potential depends on the 
$y$-component of momentum through the parameters 
$\k_\pm$, 
which are given by 
\begin{equation}
\k_\pm 
= 
\sqrt{(p_2 \pm e B \l)^2 + m^2}
,\end{equation}
and determine the asymptotic strength of the potential as 
$x \to \mp \infty$. 
The dimensionless parameter 
$\g$
is given by 
\begin{equation}
\g = \sqrt{(e B \l^2)^2 + \tfrac{1}{4}}
.\end{equation}
In the limit 
$\l \to \infty$, 
the potential becomes that of the uniform magnetic field in
Eq.~\eqref{eq:VSHO}, 
provided that the coordinate is restricted to
$|x| / \l \ll 1$. 
When 
$| x | \gtrsim \l$, 
by contrast, 
the potential does not correspond to a uniform field. 
Consequently, 
integrated quantities will not strictly reduce to uniform-field results as 
$\l \to \infty$.

The effective one-dimensional Schr\"odinger equation becomes that of Gauss's hypergeometric function 
$w(\x)$
when written in terms of the coordinate variable%
~\cite{MorseFeshbach}
\begin{equation}
\x = \frac{1}{2} \left(1 + \tanh \frac{x}{\l} \right)
\label{eq:xi}
,\end{equation} 
and the wavefunctions are additionally written in the factorized form
$\Psi(x) = \x^\a (1-\x)^\be w(\x)$, 
with the powers determined by
\begin{equation}
\a = \frac{\l}{2} \sqrt{\k_+^2 + E}
, \quad \text{and} \quad
\be = \frac{\l}{2} \sqrt{\k_-^2 + E}
.\end{equation}
For later convenience, 
the dimensionless parameter 
$z$
is defined to be their sum
$z = \a + \be$. 
The solution to the Schr\"odinger equation that is regular as 
$x \to - \infty$
(corresponding to $\x = 0$)
is
\begin{equation}
\Psi_1(x)
=
\x^\a (1- \x)^\be \, {}_2 F_1 \left(z{+}\tfrac{1}{2}{+}\g , z{+}\tfrac{1}{2} {-}\g, 1{+}2 \a | \x\right)
\label{eq:psi1}
.\end{equation}
On the other hand, 
when 
$x \to + \infty$
(corresponding to $\x = 1$), 
the regular solution has the form
\begin{equation}
\Psi_3(x)
=
\x^\a (1-\x)^\be \, {}_2 F_1  \left(z{+}\tfrac{1}{2}{+}\g , z{+}\tfrac{1}{2} {-}\g, 1{+}2 \be | 1{-}\x\right)
\label{eq:psi3}
.\end{equation}
The numbering convention for 
$\Psi(x)$
corresponds to Kummer's solutions 
$w(\x)$
of Gauss's differential equation%
~\cite{NIST:DLMF}. 
From these two solutions, 
we have the resolvent form of the coincident Green's function%
~\cite{Cangemi:1995ee} 
\begin{equation}
G(x,x|p_2, E)
=
\frac{\Psi_1(x) \Psi_3(x)}{W}
\label{eq:GResolve}
,\end{equation}
where 
$W$
is their Wronskian 
\begin{equation}
W[\Psi_1,\Psi_3]
\equiv
\Psi'_1(x) \Psi_3(x) - \Psi_1(x) \Psi'_3(x)
=
\frac{\G\left(1{+}2\a\right)\G\left(1{+}2\be\right)}{\frac{\l}{2} \, \G\left(z{+}\tfrac{1}{2}{+}\g\right)\G\left(z{+}\tfrac{1}{2}{-}\g\right)}
\label{eq:W}
.\end{equation}
With the specification of the Green's function, 
local vacuum observables can be determined from the effective action, 
such as in  
Eq.~\eqref{eq:flocal}. 
We first restrict our attention to the integrated vacuum response, 
and carry out the explicit regularization of ultraviolet divergences using dimensional regularization.

\subsection{Integrated Free Energy and Renormalization}
\label{s:IFER}

Integrated responses to the field inhomogeneity provide a glimpse into the modified dynamics within the vacuum. 
These can be expressed economically as one-dimensional integrals that have a simplicity analogous to 
Schwinger's proper-time formulation of the homogeneous problem. 
For integrated quantities, 
the trace of the effective one-dimensional Green's function 
Eq.~\eqref{eq:GResolve}
is required. 
The integrated free-energy density, 
for example, 
we denote with a bar
\begin{equation}
\overline{\c F} 
= 
\int_{- \infty}^{+\infty} \c F (x) \,  dx
=
\frac{4 \l}{3}  Z \, \frac{B^2}{2}
+ 
\overline{\c F}^\pi
,\end{equation}
which is just the effective action difference per unit 
$(d{-}1)$-dimensional spacetime volume,
or equivalently the energy per 
$(d{-}2)$-dimensional volume. 
The first term in the integrated free-energy density is the Maxwell contribution,  
while the second term
$\overline{\c F}^\p$
is the unrenormalized charged-pion loop contribution. 

For the pion vacuum fluctuations, 
we must evaluate and renormalize the integral
\begin{equation}
\overline{\c F}^\pi
=
-
\frac{2 
}{(4\p)^{\frac{d}{2}} \G\left(\frac{d}{2}\right)}
\int_{-\infty}^{+\infty} d p_2 \int_0^\infty d E \, E^{\frac{d-2}{2}}
\int_{-\infty}^{+\infty}
dx
\Big[
G(x,x|p_2, E)
-
G_{0}(0,0|p_2, E)
\Big]
\label{eq:Trace1D}
.\end{equation}
Following%
~\cite{Cangemi:1995ee} 
with further details elaborated in 
Appendix~\ref{s:integrals}, 
the one-dimensional trace can be performed in closed form, 
leading to
\begin{multline}
\overline{\c F}^\p
=
\frac{\l^2}{4 (4\p)^{\frac{d}{2}} \G\left(\frac{d}{2}\right)}
\int_{-\infty}^{+\infty} d p_2 \int_0^\infty d E \, E^{\frac{d-2}{2}}
\Bigg\{
\left( \frac{1}{\a} {+} \frac{1}{\be} \right)
\left[
\psi\left(z{+}\tfrac{1}{2}{+}\g\right)
+
\psi\left(z{+}\tfrac{1}{2}{-}\g\right)
\right]
\\
- \frac{1}{\a}
\left[ \psi(2\a{+}1) + \psi(2\a) \right]
- \frac{1}{\be}
\left[ \psi(2\be{+}1) + \psi(2\be) \right]
\Bigg\}
\label{eq:FBTrace1D}
,\end{multline}
where the potential infrared divergences cancel due to the ability to shift the 
$p_2$
integration variable within the regularized integral.%
\footnote{
There is interplay between the zero-field subtraction and the cancelation of the infrared divergence. 
In this context,  
regularization schemes employing a hard momentum cutoff become challenging to work with. 
In such schemes, 
the infrared divergence is not exactly canceled by the zero-field subtraction, 
but instead depends on the ultraviolet cutoff. 
}
In fact, 
the entire zero-field subtraction is removed in this cancelation. 
The subtracted terms that appear above are generated in performing the
$x$-integration, 
and can be rendered independent of the magnetic field due the subsequent integration over all 
$p_2$.  
It proves convenient, 
however, 
to retain this dependence and alternatively shift the integration variable 
$p_2 \to p_2 + 2 e B \l$
only in the terms containing 
$\be$, 
which produces the change
$\be \to \a$.

An efficacious change of integration variables is provided by noting the relation%
~\cite{Cangemi:1995ee}
\begin{equation}
\frac{\partial z}{\partial \c E}
= 
\frac{\l^2}{8} \left( \frac{1}{\a} + \frac{1}{\be} \right)
.\end{equation}
For the terms in the second line of 
Eq.~\eqref{eq:FBTrace1D},
one can use the analogous zero-field change of variables 
$z_0 = 2 \a$,
for which 
$\frac{\partial z_0}{\partial \c E} = \frac{\l^2}{4 \a}$.
In the resulting double integrals, 
one switches the order of integration
so that the 
$p_2$ 
integral is performed before the 
$z$
integral
(or 
$z_0$
integral in the case of terms on the second line). 
In 
$d$
dimensions, 
the 
$p_2$
integral can be performed in closed form. 
To display the result, 
we employ the two dimensionless parameters 
\begin{equation}
b = e B \l^2, 
\quad \text{and} \quad
\m = \l m
.\end{equation}
To further simplify the 
$z$
integral, 
we utilize a final change of variables to 
$z_0$, 
where
\begin{equation}
z_0 = \sqrt{z^2 - b^2},
\quad \text{so that} \quad
z = \sqrt{z_0^2 + b^2}
.\end{equation}
Carrying out this procedure, 
we arrive at the expression for the integrated free-energy density
\begin{equation}
\overline{\c F}
=
\frac{4\l}{3} 
Z \, 
\frac{B^2}{2}
+
C_d
\int_\m^\infty
dz_0 \,  (z_0^2 - \m^2)^{\frac{d-1}{2}} 
\left[ 
\psi\left(z{+}\tfrac{1}{2}{+}\g\right)
+
\psi\left(z{+}\tfrac{1}{2}{-}\g\right)
- 
\psi(z_0 {+}1)
-
\psi(z_0)
\right]
\label{eq:Fz}
,\end{equation}
where the integral is accompanied by an overall factor 
\begin{equation}
C_d 
=
\frac{\l^{1-d}}{(4 \p)^{\frac{d-1}{2}} \G \left( \frac{d+1}{2} \right)}
.\end{equation}
A salient feature of 
Eq.~\eqref{eq:Fz}
is that the vanishing of the magnetic free energy in the zero-field limit is manifest, 
due to the limiting values 
$z \to z_0$
and
$\g \to \frac{1}{2}$.

\begin{figure}[tbp]
\centering 
\includegraphics[width=.495\textwidth]{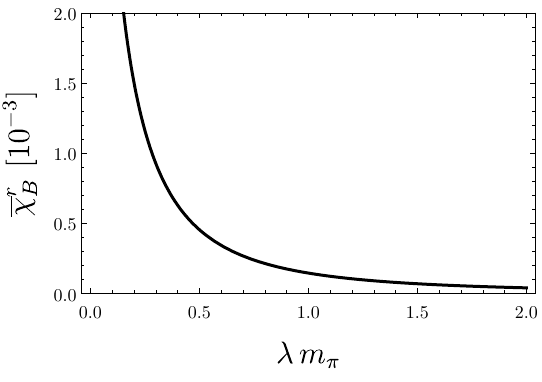}
\caption{\label{f:susc}%
Renormalized integrated magnetic susceptibility 
Eq.~\eqref{eq:chiren}
as a function of the width 
$\l$
of the magnetic field. 
The uniform-field limit is rapidly attained as a function of 
$\m = \l \, m_\p$. 
}
\end{figure}

With the above form for the integrated magnetic free-energy density, 
the sum of the integral and the contact term must be finite
using the value of the counter-term 
$h$
established in 
Eq.~\eqref{eq:h}. 
The ultraviolet divergence arises from the 
$z_0 \gg 1$
asymptotic behavior of the integrand. 
In particular,
the bracketed terms of 
Eq.~\eqref{eq:Fz}
have the leading behavior
$\sim \frac{1}{6} b^2 / z_0^{4}$, 
for which the corresponding dimensionally regulated integral is
\begin{equation}
C_d \, 
\frac{b^2}{6} 
\int_{\m}^\infty \frac{dz_0}{z_0^4} \left( z_0^2 - \m^2 \right)^{\frac{d-1}{2}}
= 
\frac{4 \l}{3 L} \, 
\frac{(eB)^2}{6(4\p)^2}
\left( \frac{1}{\e} - \g_E + \log \frac{4\p}{m^2} \right)
\label{eq:CT}
,\end{equation}
near 
$4$
dimensions. 
This divergence is indeed exactly canceled by that of the counter-term
$h$, 
leading to the renormalized result for the integrated magnetic free-energy density
\begin{equation}
\overline{\c F}
=
\frac{4 \l}{3} 
\left(
1
{-}
e^2 \, \ol \chi_B
\right)
\frac{B^2}{2}
+
\overline{\c F}^{\p,r}
\label{eq:FF}
,\end{equation}
where  
$\frac{4 \l}{3}  = \int_{-\infty}^{+\infty} B^2(x) / B^2 \, dx$
is the profile-normalizing factor.
Appearing in the second-order term of the free-energy density is an integrated magnetic susceptibility 
\begin{equation}
\ol \chi_B
=
\int dx \int dx' 
\, 
B(x) \, \chi_B(x,x') B(x')
\Bigg/
\int dx \, [B(x)]^2
.\end{equation}
This double integral is properly the susceptibility associated with the profile of the magnetic field. 
From this term, 
we can define a renormalized integrated susceptibility having the form
\begin{equation}
\ol \chi {}^r_B
(\l)
= 
\ol \chi_B (\l)
- 
\ol \chi_{B}(\infty)
\label{eq:olchir}
.\end{equation}
The subtracted term is the uniform-field susceptibility 
$\ol \chi_B (\infty) = \chi_{B}$, 
where
$\chi_B$
appears in 
Eq.~\eqref{eq:chiB}
and results from the 
$\l \to \infty$
limit of the integrated susceptibility. 
The subtraction completely removes the renormalization scale and scheme 
dependence.%
\footnote{
This subtraction is analogous to the definition of the renormalized 
magnetic susceptibility at finite temperature
$\chi_B^r(\be) = \chi_B(\be) - \chi_B(\infty)$%
~\cite{Bali:2014kia,Bali:2020bcn}.
These renormalized quantities encompass physics akin to the Casimir effect. 
In the finite temperature case, 
the boundary conditions are in the Euclidean time direction. 
For the inhomogeneous magnetic field, 
the charged-particle vacuum fluctuations respond to the finite width of the magnetic field. 
} 
The renormalized integrated magnetic susceptibility is hence a physical observable, 
for which we obtain the formula
\begin{equation}
\ol \chi_B^r(\l)
=
- \frac{4}{(4\p)^2}
\int_\m^\infty \frac{dz_0}{z_0}
\left( z_0^2 - \m^2 \right)^{\frac{3}{2}}
\left(
\psi'\left(z_0\right)
- 
\frac{1}{z_0}
-
\frac{1}{2 z_0^2}
- 
\frac{1}{6 z_0^3}
\right)
\label{eq:chiren}
,\end{equation}
where
$\psi'(z) = \frac{d}{dz} \psi(z)$
is the derivative of the digamma function.
The behavior of the renormalized integrated magnetic susceptibility is shown in 
Figure~\ref{f:susc}
as a function of the width of the magnetic field.
By construction, 
the renormalized magnetic susceptibility vanishes for large 
$\l$. 
The asymptotic behavior of 
Eq.~\eqref{eq:chiren}
shows that 
$\ol \chi^r_B 
\propto \m^{-2}$
near the uniform-field limit. 
For this reason, 
the uniform-field limit is rapidly attained as a function of 
$\m = \l \, m_\p$. 
In the narrow-width limit
(or equivalently the chiral limit), 
the renormalized magnetic susceptibility diverges
$\ol \chi^r_B 
\propto - \log \m$.

The renormalized matter contribution
$\overline{\c F}^{\p,r}$ 
is the remaining term in the integrated free-energy density in 
Eq.~\eqref{eq:FF}, 
and at next-to-leading order is given by
\begin{multline}
\overline{\c F}^{\p,r}
=
\frac{8/3}{(4\p)^2 \l^3
}
\int_\m^\infty dz_0
\left( z_0^2 - \m^2 \right)^{\frac{3}{2}}
\Bigg[
\psi\left(z{+}\tfrac{1}{2}{+}\g\right)
+
\psi\left(z{+}\tfrac{1}{2}{-}\g\right)
\\
- 
\psi(z_0 {+}1)
-
\psi(z_0)
-
\frac{b^2}{z_0}
\left(
\psi'\left(z_0\right)
- 
\frac{1}{z_0}
-
\frac{1}{2 z_0^2}
\right)
\Bigg]
\label{eq:Frenorm}
.\end{multline} 
When expressed in terms of 
$\eta = e B / m^2$
and 
$\m$, 
one sees that the renormalized integrated free-energy density is proportional to 
$m^4$ 
and appropriately linear in the width 
$\l$
of the magnetic field.

\subsection{Chiral Condensate and Renormalized Magnetization}
\label{s:integrated}

\begin{figure}[tbp]
\centering 
\includegraphics[width=.495\textwidth]{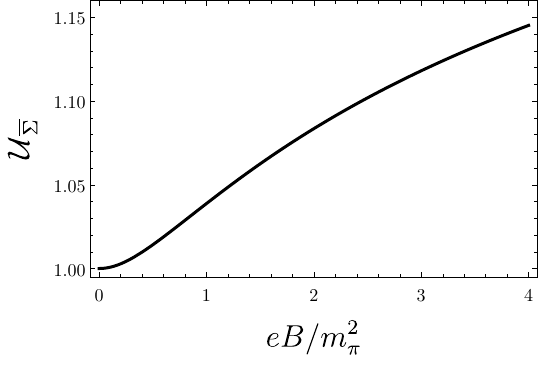}
\hfill
\includegraphics[width=.495\textwidth]{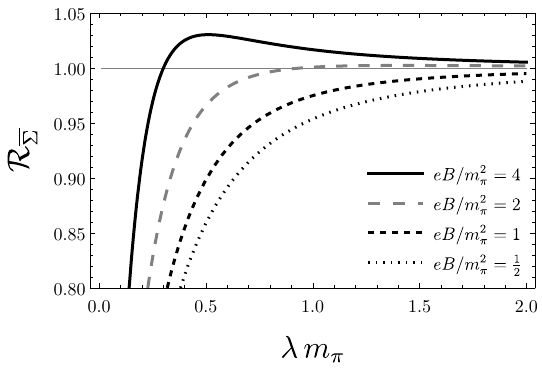}
\caption{\label{f:bulkcond}%
Modification of the chiral condensate in an inhomogeneous magnetic field. 
On the left, 
the ratio 
$\c U_{\, \ol \S}$
in 
Eq.~\eqref{eq:Ucond}
is plotted as a function of the magnetic field. 
This ratio compares the large-width limit
$\l \gg m_\p^{-1}$
of the integrated condensate to the uniform magnetic field result. 
While the ratio exceeds unity, 
it does not necessarily point to an enhancement of magnetic catalysis, 
because an overall factor of 
$\frac{4 \l}{3}$
has been used to normalize the ratio. 
On the right, 
the condensate ratio 
$\c R_{\ol \S}$
in 
Eq.~\eqref{eq:Rcond}
is plotted as a function of 
$\m = \l \, m_\p$
for a few values of the magnetic field strength. 
Enhancement of the integrated chiral condensate over the infinite-width value 
is possible for relatively large magnetic fields
$e B \gtrsim 3 m_\p^2$, 
in the small-width regime 
$\l \sim \frac{1}{4} m_\p^{-1}$. 
}
\end{figure}

The integrated difference%
\footnote{
Note that this is the spatial integral of the difference, not the difference of the spatial integrals. 
The latter difference is infrared divergent. 
} 
in the chiral condensate 
$\overline{\Delta \S}$
can be obtained from the spatial integral of 
Eq.~\eqref{eq:condD},
or from the quark-mass derivative of the integrated free energy in
Eq.~\eqref{eq:FF}. 
From either expression, 
we obtain the result for the integrated relative difference in the chiral condensate
\begin{equation}
\frac{\overline{\D \S}}{\S}
=
4
\int_\m^\infty 
\frac{dz_0 \left( z_0^2 - \m^2 \right)^{\frac{1}{2}}}{(4 \p F)^2 \l 
}
\Bigg[
\psi\left(z{+}\tfrac{1}{2}{+}\g\right)
+
\psi\left(z{+}\tfrac{1}{2}{-}\g\right)
- 
\psi(z_0 {+}1)
-
\psi(z_0)
\Bigg]
\label{eq:bulkcond}
.\end{equation}
The asymptotic behavior of the digamma functions demonstrates that the integral is finite in the ultraviolet. 
To investigate the effect of an inhomogeneous magnetic field on the integrated chiral condensate, 
we first consider the 
$\m \to \infty$
limit, 
with 
$\eta= e B / m^2$
held fixed. 
In this regime
$\l \gg m^{-1}$, 
there are only bound states in the spectrum of the one-dimensional problem; and, 
from 
Eq.~\eqref{eq:bulkcond},  
we find
\begin{equation}
\left(
\frac{\overline{\D \S}}{\S}
\right)_{\l \gg \frac{1}{m}}
=
\frac{4 m^2 \l}{(4 \p F)^2
} 
\int_1^\infty 
dz_0 \left( z_0^2 - 1 \right)^{\frac{1}{2}}
\Bigg[
\psi\left(\frac{z_0^2}{2\eta} {+} \frac{1}{2} \right)
-
\log \frac{z_0^2}{2 \eta}
\Bigg]
\label{eq:bulkcondbig}
.\end{equation}
The modification to the integrated condensate in this limit should be compared with that in a uniform magnetic field
Eq.~\eqref{eq:chiralcond}. 
To compare the two, 
we take the ratio
\begin{equation}
\c U_{\, \ol \S}
=
\frac{\qquad \left( \overline{\D \S} \, \right)_{\l \gg \frac{1}{m}} \phantom{m}}
{\frac{4\l}{3} \ \big( \S_B {-} \S \big)_\text{uniform}}
\label{eq:Ucond}
,\end{equation}
which includes the profile-normalizing factor of 
$\frac{4 \l}{3}$,
so that the ratio becomes unity in the limit of vanishing magnetic field.  
As the 
$x$-integral appearing in the expression for the integrated condensate has been performed prior to taking 
$\l \gg m^{-1}$, 
the ratio
$\c U_{\, \ol \S}$
deviates from unity, 
as shown in 
Fig.~\ref{f:bulkcond}. 
The uniform-field limit is only attained in the region of space where 
$|x| \ll \l$. 
The large width
$\m \to \infty$
behavior can be produced,
however, 
by replacing the uniform magnetic field 
$B$ 
with the profile function 
$B(x)$
in 
Eq.~\eqref{eq:Bsech}, 
and averaging over all 
$x$. 
For the modification to the condensate
Eq.~\eqref{eq:chiralcond}, 
one has
\begin{equation}
\left(\frac{\overline{\D \S} \, }{\S}\right)_{\l \gg \frac{1}{m}}
=
- \frac{m^2}{(4 \p F)^2}
\int_{-\infty}^\infty
\eta(x) \, \c I \Big( \eta^{-1}(x) \Big)
dx
,\end{equation}
with the locally constant approximation
$\eta(x) = e B(x) / m^2$. 
This relation can be verified by numerical integration, 
or by taking the asymptotic expansion of the integrand in powers of 
$B$
and comparing with the corresponding series expansion of 
Eq.~\eqref{eq:bulkcondbig}.

Additionally, 
Fig.~\ref{f:bulkcond}
explores the modification of the integrated chiral condensate as a function of the width
$\l$,
for a few values of the magnetic field. 
This is accomplished by plotting the ratio 
\begin{equation}
\c R_{ \ol \S \,  }
\equiv
\frac{\quad \overline{\D \S}  \qquad}
{\quad \Big( \overline{\D \S} \, \Big)_{\l \gg \frac{1}{m}} }
\label{eq:Rcond}
,\end{equation}
as a function of 
$\m$. 
By definition, 
the ratio of the condensate modification 
$\c R_{\ol \S}$
tends to unity as 
$\l \to \infty$, 
however, 
the approach is seen to be from below for 
$e B \lesssim 3 m_\p^2$, 
and from above for 
$e B \gtrsim 3 m_\p^2$.
When the width
$\lambda$
is a fraction of the pion Compton wavelength, 
the integrated condensate can be enhanced over the infinite-width value as the magnetic field strength increases.
The strengths required for 
$\sim 10\%$
enhancement, 
namely
$e B \sim 10 \, m_\p^2$,
are nearly beyond the reach of chiral perturbation theory, 
however,
which requires
$e B / (4 \p F_\p)^2 \ll 1$
to neglect the next-order corrections. 
Note that for small widths  
$\l \ll m_\p^{-1}$, 
the integrated chiral condensate appearing in
Eq.~\eqref{eq:bulkcond}
is proportional to 
$\l^{-1}$. 
The precipitous drop in the ratio 
$\c R_{\ol \S}$
is proportional to 
$\l^{-2}$
for small 
$\l$, 
which owes to the additional linear 
dependence on 
$\l$
from the denominator of
Eq.~\eqref{eq:Rcond}.

\begin{figure}[tbp]
\centering 
\includegraphics[width=.495\textwidth]{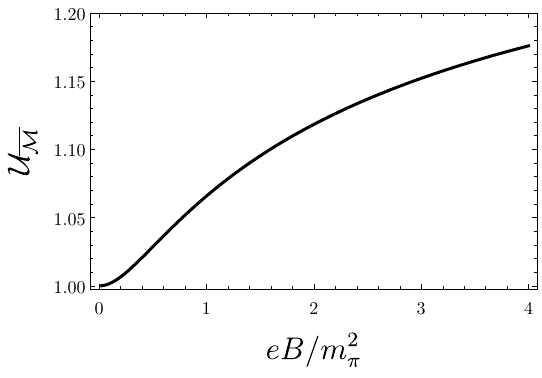}
\hfill
\includegraphics[width=.495\textwidth]{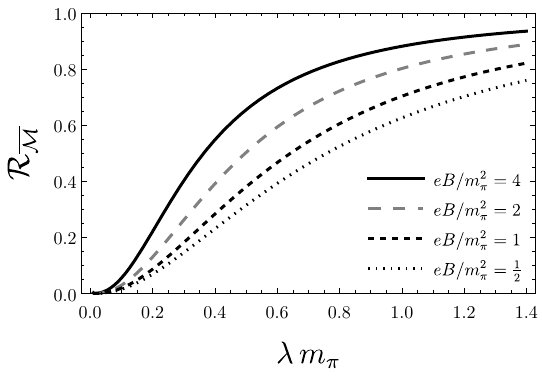}
\caption{\label{f:bulkmag}%
Modification of the renormalized magnetization in an inhomogeneous magnetic field. 
On the left, 
the ratio 
$\c U_{\overline{\c M}}$
in 
Eq.~\eqref{eq:Umag}
is plotted as a function of the magnetic field. 
This ratio compares the large-width limit
$\l \gg m_\p^{-1}$
of the integrated magnetization to the uniform magnetic-field result. 
On the right, 
the ratio 
$\c R_{\overline{\c M}}$
in 
Eq.~\eqref{eq:Rmag}
is plotted as a function of 
$\m = \l \, m_\p$,
for a few values of the magnetic field strength. 
The infinite-width limit of the integrated magnetization is more rapidly attained for larger field strengths. 
}
\end{figure}

From the integrated free-energy density
Eq.~\eqref{eq:Frenorm}, 
the response to the field strength 
$B$
still encodes the magnetization of the system, 
albeit from a non-uniform response averaged over all space.%
\footnote{
The spatially integrated vacuum current vanishes due to parity. 
} 
Using 
Eq.~\eqref{eq:M}, 
the renormalized integrated magnetization is given by the formula
\begin{multline}
\overline{\c M} {}^r
=
-
\frac{8/3 \, e \, b}{(4\p)^2 \l}
\int_\m^\infty dz_0
\left( z_0^2 {-} \m^2 \right)^{\frac{3}{2}}
\Bigg[
\left( \frac{1}{z} {+} \frac{1}{\g} \right) 
\psi'\left(z{+}\tfrac{1}{2}{+}\g\right)
+
\left( \frac{1}{z} {-} \frac{1}{\g} \right) 
\psi'\left(z{+}\tfrac{1}{2}{-}\g\right)
\\
- \frac{2}{z_0}
\left(
\psi'(z_0) - \frac{1}{z_0} - \frac{1}{2 z_0^2}
\right)
\Bigg]
\label{eq:bulkM}
.\end{multline} 
The bracketed terms have 
asymptotic behavior 
$\propto z_0^{-8}$
for 
$z_0 \gg 1$, 
which leads to better convergence in the ultraviolet compared to the free-energy density. 
We investigate the integrated magnetization analogously to the integrated condensate. 
First, 
we form the ratio of the integrated magnetization in an inhomogeneous field of large width
compared to that of a uniform magnetic field 
\begin{equation}
\c U_{\overline{\c M}}
=
\frac{\overline{\c M} {}^r_{\l \gg \frac{1}{m}}}{\frac{32 \l}{35} \,\c M^r_\text{uniform}}
\label{eq:Umag}
,\end{equation}
where 
$\frac{32 \l}{35} = \int_{-\infty}^{+\infty} B^4(x) / B^4  \, dx$. 
This ratio is plotted in 
Fig.~\ref{f:bulkmag}
as a function of the magnetic field strength, 
which shows deviation from unity. 
As above, 
the 
$\m \to \infty$
limit does not result in the uniform-field result due to the lack in uniformity of convergence in 
$x$. 
Instead, 
the ratio starts at unity for small fields by the choice of normalization in 
Eq.~\eqref{eq:Umag}, 
and the behavior can be accounted for by replacing 
$B$ 
in the uniform-field answer
Eq.~\eqref{eq:FB} 
with the profile 
$B(x)$, 
and subsequently averaging over all 
$x$. 
Note the replacement 
$B \to B(x)$
must be carried out for the magnetic free-energy density, 
so that differentiation with respect to the field strength parameter
$B$ 
will then appropriately yield the integrated magnetization.

Additionally shown in   
Fig.~\ref{f:bulkmag}
is the ratio of the  renormalized integrated magnetization to its large-width limit
\begin{equation}
\c R_{\overline{\c M}}
\equiv
\frac{\overline{\c M} {}^r}
{\quad \, \overline{\c M} {}^r_{\l \gg \frac{1}{m}}}
\label{eq:Rmag}
,\end{equation}
which is plotted as a function of 
$\m$, 
for a few values of the magnetic field strength. 
By construction, 
the ratio becomes unity for 
$\l \gg m_\p^{-1}$, 
with this limit more rapidly attained for larger field strengths. 
In general, 
the renormalized magnetization is depleted compared to its large-width value. 
For small 
$\l$, 
the integrated magnetization in 
Eq.~\eqref{eq:bulkM}
is proportional to 
$\l^4$. 
Accordingly 
the integrated magnetization ratio 
$\c R_{\overline{\c M}}$
vanishes 
$\propto \l^3$
due to the linear dependence on 
$\l$
from the denominator.

While the formulas for integrated vacuum observables are relatively simple, 
the integration over all space tends to obscure the physical effects. 
This is due to the lack of uniform convergence as a function of the 
$x$ 
coordinate. 
The physics is directly accessible from local quantities, 
albeit at the cost of more complicated integrals that are evaluated below.

\section{Inhomogeneous Magnetic Field: Local Quantities}
\label{sec:local}

\subsection{Free-Energy Density}
\label{sec:freee}

The general expression for a representation of the local magnetic free-energy density 
$\c F(x)$
is given in 
Eq.~\eqref{eq:flocal}, 
and depends on the Green's function 
Eq.~\eqref{eq:GResolve}. 
The counter-term
$h$, 
moreover, 
has been determined in dimensional regularization, 
from which we infer that the integral must contain a divergence proportional to the local energy density of the magnetic field
$\frac{1}{2} B^2(x)$. 
With forethought, 
the divergence can be exhibited and consequently canceled, 
leading to integrals that are ultimately amenable to numerical evaluation.

As divergences arise from multiple asymptotic regions in the 
$p_2$--$E$
plane, 
a change of variables becomes preferable to isolate these divergences.
The new variables are chosen to be
$z_0 = \sqrt{z^2 {-} b^2}$
with 
$z = \a + \be$, 
exactly as above, 
and
\begin{equation} 
\D = 
\frac{\d}{ %
\sqrt{1 {-} \frac{\m^2}{z_0^2}}},
\quad \text{with} \quad
\d = \frac{\a - \be}{b} 
,\end{equation}
where 
$\text{sign} (\D) 
= 
\text{sign} (p_2)$. 
A salient feature of this variable transformation is that
$|p_2| \to \infty$
corresponds to 
$z_0 \to \infty$
with 
$\D$ 
finite,
which is additionally true as 
$E \to \infty$
independent of the size of
$p_2$. 
The ultraviolet divergences are consequently restricted to the asymptotic region in the 
$z_0$--$\D$ plane
where
$z_0 \gg 1$.
To facilitate evaluation of the integrand,  
the Green's function can be re-expressed by performing one of Kummer's transformations
\begin{equation}
{}_2 F_1 (a,b,c|z)
= 
(1-z)^{c-a-b} \
{}_2 F_1(c{-}a,c{-}b,c|z)
,\end{equation}
on each hypergeometric function, 
which results in the product of wavefunctions being converted to the product of only hypergeometric functions
\begin{equation}
\Psi_1(x) \Psi_3(x)
=
{}_2 F_1 \big(\tfrac{1}{2}{+}b
\d{+}\g,\tfrac{1}{2}{+}b
\d{-}\g, 1 {+} z {+} b
\d |\x\big)
\, {}_2 F_1 \big(\tfrac{1}{2}{-}b
\d{+}\g,\tfrac{1}{2}{-}b
\d{-}\g, 1 {+} z {-} b
\d  |1{-}\x\big)
,\end{equation}
where for brevity
we subsequently use 
$z = \sqrt{z_0^2 {+} b^2}$
and
$\d = 
\sqrt{1 {-} \frac{\m^2}{z_0^2}} \, \D$
as abbreviations.

The local free-energy density must be an even function of 
$x$. 
In terms of the coordinate variable
$\x$
defined in
Eq.~\eqref{eq:xi}, 
this requires symmetry under the reflection 
$\x \to 1{-}\x$. 
This symmetry can be made manifest by separating the
$\D$
integral into contributions from positive and negative regions, 
and making the substitution 
$\D \to -\D$
in the latter. 
Combining the functional identities with the change of variables,  
the representation of the local magnetic free-energy density 
Eq.~\eqref{eq:flocal}
can be written as
\begin{equation}
\c F(x)
=
Z \, \frac{B^2(x)}{2} 
-
\c C_d
\int_\mu^\infty dz_0 \,
(z_0^2 {-} \m^2)^{\frac{d-1}{2}}
 \int_{0}^{1} d\D
 (1 {-} \D^2)^{\frac{d-2}{2}}
\,  \c G \big(z_0,\d(\D) \big| x \big)
\label{eq:fintermed}
.\end{equation}
The dependence on the number of dimensions 
$d$
appears in the multiplicative prefactor of the double integral
\begin{equation}
\c C_d
= \frac{
4\,
\l^{-d}}{(4\p)^{\frac{d}{2}} \G \big(\frac{d}{2} \big)}
\label{eq:Cdpre}
,\end{equation}
as well as the two factors exhibited in the integrand of 
Eq.~\eqref{eq:fintermed}.  
The remaining functional dependence of the integrand is described by 
$\c G(z_0, \d | x)$, 
which is 
$d$-independent and given by
\begin{equation}
\c G(z_0,\d | x)
=
\frac{\Gamma(z{+}\frac{1}{2}{+}\g) \Gamma(z{+}\frac{1}{2}{-}\g)}
{2 z\, \Gamma(z{+} b
\d) \Gamma(z{-}b
\d)}
\Big[
\Psi_1(x) \Psi_3(x) 
{+} 
\Psi_1({-}x) \Psi_3({-}x)
\Big]
-
1
\label{eq:Gzdxi}
,\end{equation}
where the subtraction term has first been simplified due to the ability to shift the 
$p_2$
integration variable in individual terms of the regulated integral
Eq.~\eqref{eq:flocal}. 
The above function is appropriately even in 
$x$, 
as is the local magnetic free-energy density.

The ultraviolet behavior of the double integral requires scrutiny. 
The logarithmic divergence must be canceled by the counter-term 
$h$
in 
Eq.~\eqref{eq:h};
and, 
on account of 
Eq.~\eqref{eq:fintermed}, 
it can be rewritten in the efficacious form
\begin{equation}
h
=
-
\l^4 \, \c C_d
\int_\m^{\infty} \frac{dz_0}{
4 \,
z_0^4} 
(z_0^2 {-} \m^2)^{\frac{d-1}{2}} \int_0^1 d\D (1{-}\D^2)^{\frac{d-2}{2}}
+
\frac{\ol h}{3(4\p)^2}
\label{eq:hlocal}
,\end{equation}
near 
$d{=}4$
dimensions. 
After accounting for the counter-term in this form, 
the local magnetic free-energy density becomes
\begin{equation}
\c F(x)
=
\frac{1}{2} \left[1 + \frac{e^2 \, \ol h}{3 (4 \p)^2} \right] B^2(x)
-
\c C_d
\int_\mu^\infty dz_0 \,
(z_0^2 {-} \m^2)^{\frac{d-1}{2}}
\int_{0}^{1} d\D
(1 {-} \D^2)^{\frac{d-2}{2}}
\,
\overline{\c G} 
\big(z_0, \d(\D) \big| x \big)
\label{eq:fintermed2}
,\end{equation}
where the modified function 
$\overline{\c G} (z_0, \d | x )$
is defined by
\begin{equation}
\overline{\c G} (z_0, \d | x )
=
\c G (z_0, \d | x )
+
\frac{2 b^2}{z_0^4} \x^2 (1{-}\x)^2
\label{eq:Gbar}
,\end{equation}
for which the additional term removes the logarithmic divergence from the double integral.
In Eq.~\eqref{eq:fintermed2},
however, 
we cannot merely set 
$d{=}4$, 
because there is still a divergence present that requires dimensional regularization. 
The subtraction of the zero-field Green's function in 
$d{=}4$
dimensions,
which appears as the 
$x$-independent piece of the function 
$\c G(z_0,\d |x)$, 
removes the quartic divergence of the  
$z_0$
integral.
The remaining divergence can be exhibited by writing out the 
$z_0 \gg 1$
behavior of the integrand in 
Eq.~\eqref{eq:fintermed2}. 
Taken in 
$d{=}4$
dimensions 
with the integral over 
$\D$
performed, 
we would have 
\begin{equation}
- \frac{4}{\l^4 (4\p)^2}
\int_\m^\infty 
dz_0
\left[
\frac{8}{15} b^2 \x (1{-}\x)
\, z_0
+
\c O(z_0^{-3})
\right]
\label{eq:powerdiv}
,\end{equation}
exposing a quadratic divergence. 
Such a power-law divergence, 
however,  
is absent in dimensional regularization. 
Schematically, 
the offending term arises from the leading asymptotic behavior
$\propto z_0^{-2}$
of 
$\c G \big(z_0,\d(\D) \big| x \big)$
in
Eq.~\eqref{eq:Gzdxi}, 
which in $d$ dimensions then produces%
\footnote{
This is analogous to a scalar tadpole loop integral, 
which in $d$-dimensional momentum space is given by the expression
\begin{equation}
\notag
\int \frac{d^d k}{(2\p)^d} \frac{1}{k^2 + m^2}
=
\frac{m^2}{(4\p)^2}
\left( \frac{4\p}{m^2} \right)^{\e}
\G(-1{+}\e)
,
\quad \text{for} \quad 
\e > 1
.\end{equation}
Having evaluated the integral in strictly 
$d < 2$
dimensions, 
the result then allows analytic continuation to $d$
near 
four dimensions, 
for which 
$0 < \e \ll 1$.
The integral is quadratically divergent in 
$d {=} 4$
dimensions. 
} 
\begin{equation}
\int_\m^\infty  \frac{dz_0}{z_0^2} (z_0^2 - \m^2)^{\frac{3}{2} - \e}
\propto 
\G(-1{+}\e),
\quad \text{for} \quad
\e > 1
\label{eq:power}
.\end{equation}
The result is analytically continued around the pole at 
$\e = 1$
to enable expansion about the pole at
$\e = 0$. 
Accordingly, 
dimensional regularization automatically renormalizes the power-law divergence.%
\footnote{
Note that in a different regularization scheme, 
one might be required to add a gauge-dependent counter-term 
$\propto \left[ e A_\m(x) \right]^2$
to remove the power-law divergence exhibited in 
Eq.~\eqref{eq:powerdiv}.
}
When one additionally accounts for the sub-leading asymptotic behavior
of the integrand of Eq.~\eqref{eq:fintermed2}, 
these terms produce a pole at 
$\e = 0$ 
directly through the function
$\G(\e)$, 
but the pole is exactly canceled by the 
$\e \ll 1$
expansion of the analytically continued result in 
Eq.~\eqref{eq:power}. 
There is no remaining logarithmic divergence; 
it has been completely canceled by the counter-term.

In the present situation, 
we cannot perform the complete integral as a function of 
$\e$
to enable the analytic continuation. 
Instead, 
we add and subtract the asymptotic behavior of the integrand, 
and carry out the analytic continuation on the added term. 
The subtracted term accordingly removes the troublesome ultraviolet behavior
and allows for the limit 
$d \to 4$
to be taken in the integrand. 
In Eq.~\eqref{eq:fintermed2}, 
the function 
$\overline{\c G} \big(z_0, \d(\D) \big| x \big)$
has the 
$z_0 \gg 1$
asymptotic behavior described by 
\begin{multline}
g(z_0,\D | x)
=
\frac{2 b^2 }{z_0^2} \x (1{-}\x)
\Bigg(
1 {-} 3\D^2
+ 
\frac{1}{z_0^2}
\Bigg\{
\big(1 {-} 5 \D^2 \big) 
\Big[ 1 {-} 5 \x (1{-}\x) \Big]
+
3 \D^2 \m^2
\\
-
b^2 
\Big[
1 
{-}
8 \D^2 
{+} 
7 \D^4
{-} 
\x (1{-}\x) 
\Big(
3
{-} 
30 \D^2
{+}
35 \D^4
\Big)
\Big]
\Bigg\}
\Bigg)
\label{eq:littleg}
,\end{multline}
up to terms of
$\c O(z_0^{-5})$
that give ultraviolet finite contributions to 
Eq.~\eqref{eq:fintermed2}.  
Using dimensional regularization, 
we have the finite result from the added term
\begin{eqnarray}
g (x)
&\equiv&
\int_\m^\infty dz_0 \, (z_0^2 - \m^2)^{\frac{d-1}{2}} \int_0^1 d\D (1-\D^2)^{\frac{d-2}{2}} g(z_0,\D| x)
\notag \\
&=&
-\frac{4 b^2}{105} \x (1 {-} \x)
\Big\{
7 - 35 \x (1{-}\x)
-
2 b^2 \Big[ 2 {-} 3 \x (1{-}\x) \Big]
\Big\}
,\end{eqnarray}
after analytic continuation
and subsequently taking the 
$\e \to 0$
limit. 
The expression for this representation of the local free-energy density takes the form
\begin{equation}
\c F(x) 
= 
\frac{1}{2} \left[ 1+ \frac{e^2 \, \ol h}{3 (4\p)^2} \right] 
B^2(x) + \c F_B(x)
\label{eq:flocal_general}
,\end{equation}
where the local 
(partially renormalized)
magnetic free-energy density is given by
\begin{multline}
\c F_B(x) 
= 
-
\frac{4}{\l^4 (4\p)^2}
\Bigg\{
g (x)
+
\int_\m^\infty dz_0 \, (z_0^2 {-} \m^2)^{\frac{3}{2}}
\int_0^1 d\D (1{-}\D^2)
\Big[
\overline{\c G} \big(z_0, \d(\D) \big| x \big) 
- 
g(z_0,\D|x)
\Big]
\Bigg\}
\label{eq:fBlocal}
.\end{multline}
The integrand has been evaluated in 
$d{=}4$
dimensions, 
but the double integral is now demonstrably finite in the ultraviolet. 
Recovery of the integrated free-energy density from integrating 
this expression for the local free energy is discussed in 
Appendix~\ref{s:integrate}.

\begin{figure}[tbp]
\centering 
\includegraphics[width=.495\textwidth]{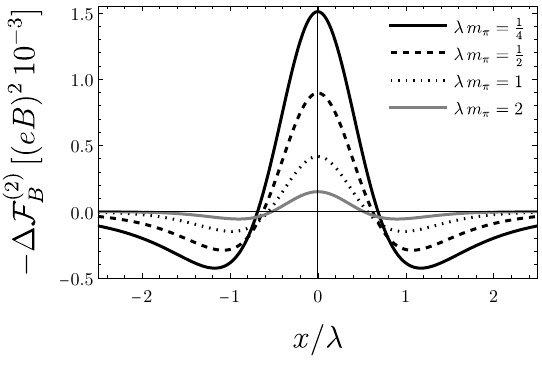}
\caption{\label{f:quadratic}%
Deviation of the quadratic response of the free-energy density from the energy density in the magnetic field. 
The negative of the difference 
$\D \c F_B^{(2)}$
in 
Eq.~\eqref{eq:DeltaF2}
is plotted as a function of 
$x$
in units of 
$(eB)^2$.
Because the renormalized free energy in a uniform magnetic field contains no corresponding 
$\c O( B^2)$ 
contribution,
the residual quadratic term provides a signature of the nonlocal response of charged-pion fluctuations to the magnetic-field profile.
}
\end{figure}

To complete the renormalization of the local magnetic free-energy density, 
we add and subtract the residual
$\c O(B^2)$ 
contribution. 
For an inhomogeneous system,
this residual contribution is a projection of the nonlocal magnetic susceptibility%
\footnote{
As we have removed the local Maxwell contribution, 
this is a renormalized magnetic susceptibility, 
namely
$\chi^r_B(x,x') = \chi_B (x,x') + \delta(x {-} x')   \frac{\ol h}{3 (4\p)^2}$.
} 
\begin{equation}
\c F_B^{(2)} (x)
=
-
\frac{1}{2} e B(x) \int_{-\infty}^{+\infty}  \chi^r_B(x,x') \, e B(x') \, dx'
.\end{equation} 
When integrated over all space, 
one arrives at the renormalized integrated susceptibility that is introduced in 
Eq.~\eqref{eq:olchir}
\begin{equation}
\int_{-\infty}^{+\infty} \c F_B^{(2)} (x) \, dx
=
-
\frac{4 \l}{3} \, \overline{\chi}_B^r (\l) \, \frac{(eB)^2}{2}  
.\end{equation}
The renormalized magnetic free-energy density is then defined to be
\begin{equation}
\c F^r_B(x) = \c F_B(x) - \c F_B^{(2)}(x)
.\end{equation} 
As 
$\c F_B(x)$
is entirely due to charged-pion vacuum fluctuations, 
this further separation of the 
$\c O(B^2)$
contribution is a matter of convention inherited from homogeneous systems, 
as well as the corresponding integrated observables in inhomogeneous systems.

The residual second-order contribution nonetheless provides a useful probe of the nonlocal response to the magnetic-field profile.
Such a contribution is absent in the uniform field result, 
where the renormalized free energy begins at 
$\c O(B^4)$.  
The appearance of a finite 
$\c O(B^2)$ 
contribution in the inhomogeneous case therefore reflects the response of charged-pion fluctuations to the spatial structure of the magnetic field.
In particular, 
the difference 
\begin{equation}
\D \c F_B^{(2)}
(x)
=
{\c F_B^{(2)}(x)}
-
\left( -
\frac{1}{2} \, \overline{\chi}_B^r (\l) \, e^2 B^2(x)
\right)
\label{eq:DeltaF2}
,\end{equation}
compares the exact quadratic response with a local reference profile normalized to produce the same 
integrated quadratic response, 
so that 
$\int_{-\infty}^{+\infty} \D \c F_B^{(2)}(x)  \, dx = 0$. 
The negative of this difference is plotted in
Fig.~\ref{f:quadratic}
and shows the deviation of the quadratic response from the profile of the magnetic energy density. 
While the absolute size and extent of the deviation is reduced with increasing profile width 
$\l$, 
the fractional deviation remains substantial.  
This is consistent with the absence of a corresponding 
$\c O(B^2)$ 
contribution in the uniform-field limit, 
indicating that the residual quadratic response is intrinsically tied to the spatial variation of the magnetic field.

Having exhibited this 
$\c O(B^2)$
feature of the local free-energy density obtained from the exact resolvent form of the Green's 
function, 
we turn to the chiral condensate and vacuum current. 
In contrast to the local free-energy density, 
which is only defined up to a total derivative, 
these local quantities are uniquely defined because they are obtained from functional differentiation 
of the effective action with respect to local sources. 
Consequently, 
they admit an unambiguous local interpretation, 
whereas only the integrated free energy is physically meaningful.

\subsection{Chiral Condensate}

The local chiral condensate can be determined from 
Eq.~\eqref{eq:condD}. 
As with the local free-energy density,  
the variable transformation from 
$p_2$, $E$
to 
$z_0$, $\D$
proves efficient for numerical evaluation. 
We find the expression for the relative difference in the chiral condensate%
\footnote{
One can alternatively start with the renormalized free-energy density in 
Eq.~\eqref{eq:fBlocal}
and take the pion mass-squared derivative. 
The local chiral condensate in Eq.~\eqref{eq:condlocal2}
is indeed reproduced, 
provided one accounts for the pion-mass dependence of the counter-term 
$\ol h$.
}
\begin{equation}
\frac{\S(x) {-} \S}{\S}
=
-
\frac{\l^2 \, \c C_d}{F^2} 
\, \frac{d{-}2}{2}
\int_\mu^\infty dz_0
\left( z_0^2{-}\m^2 \right)^{\frac{d-3}{2}}
\int_0^1 d\D (1 {-} \D^2)^{\frac{d-4}{2}}
\, \c G \big( z_0, \d(\D) \big| x \big)
\label{eq:condlocal}
,\end{equation}
where the $d$-dimensional pre-factor 
$\c C_d$
is that defined in 
Eq.~\eqref{eq:Cdpre}, 
and the function 
$\c G \big( z_0, \d \big| x \big)$
appears in 
Eq.~\eqref{eq:Gzdxi}.  
One must be careful in taking the 
$d \to 4$
limit of the expression for the  local condensate in order to account for all finite terms.
In contrast to the free-energy density,  
the 
$z_0 \gg 1$
asymptotic behavior of the integrand for the condensate requires only the leading term of 
Eq.~\eqref{eq:littleg}
\begin{equation}
\widetilde{g} \big( z_0, \D \big| x \big)
=
\frac{2b^2}{z_0^2}(1{-}3\D^2) \x (1{-}\x)
\label{eq:braces}
.\end{equation}
The $z_0$ integral of this asymptotic term leads to a logarithmic divergence, 
which appears proportional to  
$\G(\e)$
in dimensional regularization.
The 
$\D$ integral,
however,
is proportional to 
$\e$
leading to an overall finite contribution
\begin{equation}
\widetilde{g}(x)
\equiv
\int_\mu^\infty dz_0
\left( z_0^2{-}\m^2 \right)^{\frac{d-3}{2}}
\int_0^1 d\D (1 {-} \D^2)^{\frac{d-4}{2}}
\, \, \widetilde{g} ( z_0, \D | x )
=
-
\frac{2}{3} b^2 \x (1{-}\x)
,\end{equation}
as 
$\e \to 0$. 
The final expression for the local chiral condensate is thus
\begin{equation}
\frac{\S(x) {-} \S}{\S}
=
-
\frac{4}{(4\p \l F)^2} 
\Bigg\{
\widetilde{g}(x)
+
\int_\mu^\infty dz_0
\left( z_0^2{-}\m^2 \right)^{\frac{1}{2}}
\int_0^1 d\D 
\Big[
\c G \big( z_0, \d(\D) \big| x \big) 
{-}
\widetilde{g}(z_0,\D |x)
\Big]
\Bigg\}
\label{eq:condlocal2}
,\end{equation}
for which the subtraction term of the integrand guarantees convergence of the 
$z_0$
integral in the ultraviolet.

\begin{figure}[tbp]
\centering 
\includegraphics[width=.495\textwidth]{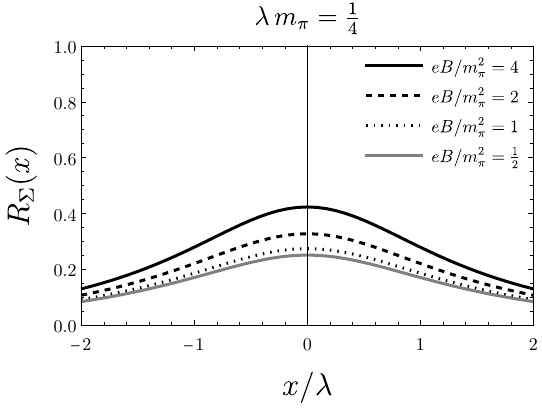}
\hfill
\includegraphics[width=.495\textwidth]{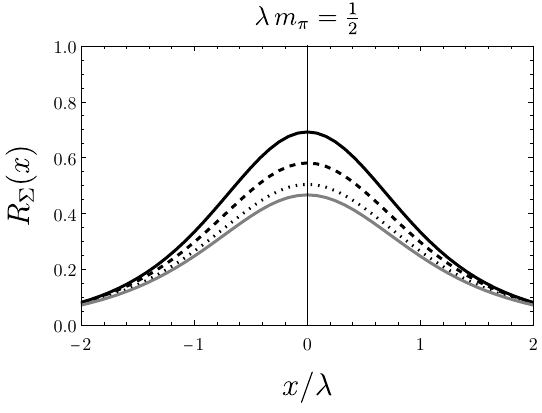}
\\
\includegraphics[width=.495\textwidth]{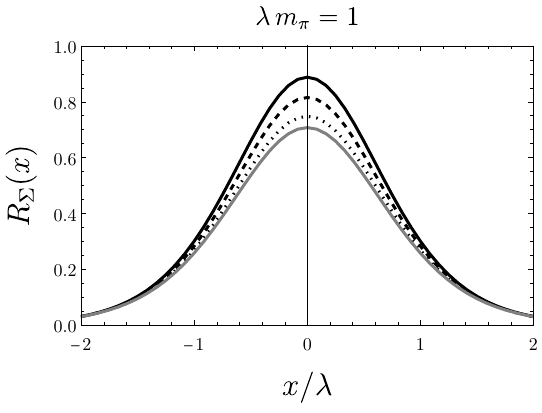}
\hfill
\includegraphics[width=.495\textwidth]{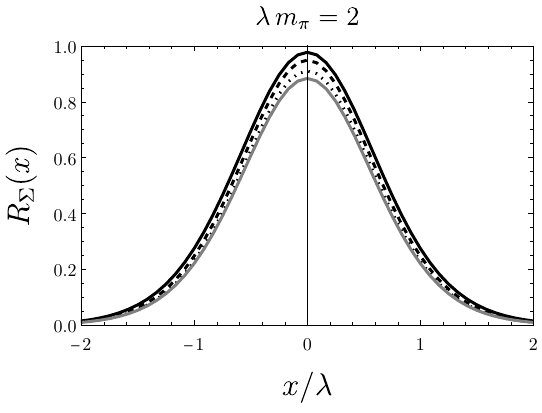}
\caption{\label{f:localcond}%
Spatial variation of the chiral condensate.
The ratio 
$R_{\S}(x)$
in 
Eq.~\eqref{eq:Rlocal}
is plotted as a function of 
$x$
for differing widths
$\l$
of the magnetic field, 
each for a few values of the magnetic field strength
$eB$.
The field-strength legend provided in the first plot applies to all four plots. 
Note that the overall widening or narrowing with respect to 
$\lambda$
has been removed by plotting the condensate as a function of 
$x / \lambda$. 
With this taken into account, 
the narrower field profile leads to comparatively larger tails of the chiral condensate. 
The narrower profile, 
moreover, 
has the most dramatic reduction of the size of the condensate compared to the corresponding 
uniform-field value. 
As the width of the field increases, 
there is progressively less magnetic-field dependence in the ratio.
This is expected as the 
$\l \to \infty$
limit for 
$x / \l$ 
near zero 
must become identical to the uniform-field condensate for all values of 
$eB$.
} 
\end{figure}

The behavior of the local chiral condensate is shown in 
Fig.~\ref{f:localcond}. 
Specifically the ratio of the modification to the condensate compared to that in a uniform magnetic field of the 
same field strength 
\begin{equation}
R_{\S}(x)
=
\frac{\phantom{uni} \S(x) - \S \phantom{unifo}}
{\phantom{uni}\big(\S_B - \S \big)_\text{uniform}}
\label{eq:Rlocal}
,\end{equation}
is plotted as a function of 
$x$. 
The figure compares this ratio's dependence on the width 
$\l$
of the magnetic field, 
each for a few values of the magnetic field strength 
$eB$.
The obvious narrowing 
(widening) 
of the profile with decreasing
(increasing)
$\lambda$
is removed by plotting the condensate as a function of the scaled coordinate
$x / \lambda$. 
The catalysis of chiral symmetry breaking is greatly reduced for magnetic fields having a narrow profile. 
A narrow-profile magnetic field, 
moreover, 
has comparatively larger tails in the local chiral condensate. 
A wide-profile magnetic field leads to the uniform-field limit, 
but not uniformly as a function of 
$x$, 
which is confirmed by the last plot shown in 
Fig.~\ref{f:localcond}.
The condensate ratio tends to unity only in the region in which 
$| x | / \l \ll 1$.

\begin{figure}[tbp]
\centering 
\includegraphics[width=.495\textwidth]{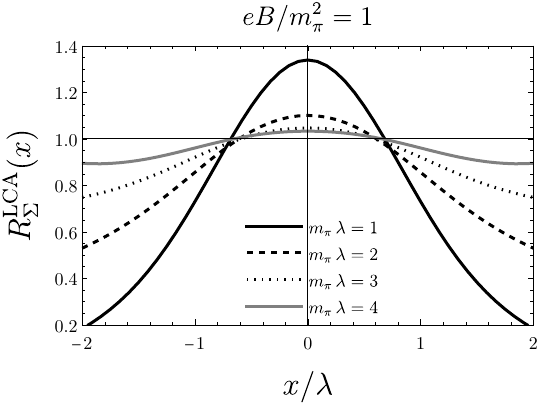}
\hfill
\includegraphics[width=.495\textwidth]{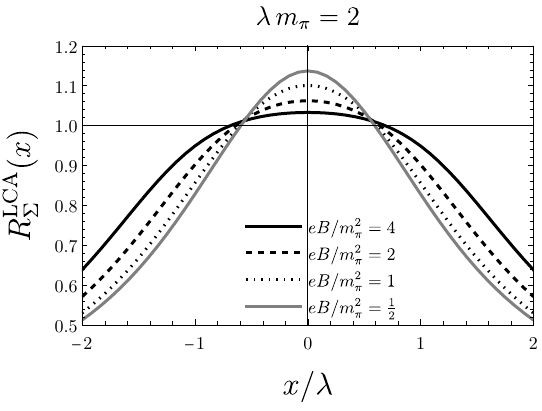}
\caption{\label{f:localcondLCA}%
Locally constant approximation to the chiral condensate.
On the left, 
the ratio 
$R^\text{LCA}_{\S}(x)$
in 
Eq.~\eqref{eq:localcondLCA}
is plotted as a function of 
$x$
using differing widths
$\l$
of the magnetic field, 
for a fixed magnetic field strength
$eB$.
The locally constant approximation describes the local condensate when the value of the ratio is near unity, 
and the approximation becomes exact in the large-width limit.  
On the right, 
the same ratio is plotted for differing magnetic field strengths, but with the same spatial width.  
The locally constant approximation is shown to improve with increasing field strength. 
}
\end{figure}

Another quantity of interest is the locally constant approximation to the spatial variation of the condensate.  
This approximation arises from replacing the field strength parameter 
$\eta = e B / m^2$
in the uniform-field result 
Eq.~\eqref{eq:chiralcond} 
with 
$\eta(x) = e B(x)/m^2$. 
This locally constant approximation can be tested against the local condensate through the ratio 
\begin{equation}
R_{\S}^{\text{LCA}}
=
\frac{\phantom{uni}\big(\S_{B(x)} - \S \big)_\text{uniform}}
{\phantom{uni} \S(x) - \S \phantom{unifo}}
\label{eq:localcondLCA}
,\end{equation}
where we denote the locally constant approximation by 
$\S_{B(x)}$. 
Locally constant approximation ratios are shown in 
Fig.~\ref{f:localcondLCA}. 
For a fixed value of the magnetic field strength
$eB$, 
the ratio is shown for a few moderate-size widths
$\l$. 
The approximation is progressively better for larger widths, 
but systematically overestimates the chiral condensate where the magnetic field is greatest,
while underestimating where the field is smallest.%
\footnote{
This overestimation and underestimation largely cancels out when integrated over all space, 
and leads to the rapid approach to the uniform-field limit shown in 
Sec.~\ref{s:integrated}
for the integrated condensate.  
Such a rapid approach is consequently not shared by the local chiral condensate. 
} 
For smaller widths, 
the locally constant approximation fails. 
This failure demonstrates that magnetic catalysis is spatially nonlocal in an inhomogeneous background. 
The condensate does not respond only to the local value of 
$B(x)$, 
but to the magnetic profile over the pion Compton wavelength. 
As the width of the magnetic profile increases, 
the response to the slowly varying field becomes adiabatic. 
In this regime,  
by contrast,
the locally constant approximation becomes exact.
The figure also contrasts the locally constant approximation at a fixed width, 
but for different values of the field strength. 
The approximation improves with increasing field strength, 
and has the same pattern of overestimation in the center and underestimation in the tails.

\subsection{Vacuum Current}

The current provides another observable that probes the vacuum response to an inhomogeneous magnetic field. 
The induced vacuum current can be obtained from the general expression in 
Eq.~\eqref{eq:currentgen}. 
Only the $y$-component of the current is non-vanishing, 
and it has two contributions: 
the source current and the charged-pion vacuum fluctuations
\begin{equation}
J_y(x)
=
-
Z\,
\frac{dB}{dx}
+ 
J_y^\p(x)
.\end{equation}
The unrenormalized vacuum fluctuations are given by a momentum-space integral over the gauge-covariant derivative
of the coincident Green's function
\begin{equation}
J_y^\p(x)
=
- \frac{4 \, e}{(4\p)^{\frac{d}{2}} \G\left(\frac{d{-}2}{2}\right)} 
\int_{-\infty}^{+\infty} dp_2 \int_0^\infty d E \,  E^{\frac{d-4}{2}}
\big[ p_2 - e A(x) \big] \,
G(x,x|p_2, E)
.\end{equation}
The integral is dimensionally regulated, 
but requires renormalization.

Once more, 
the variable transformation from 
$p_2$, $E$
to 
$z_0$, $\D$
proves useful for isolating the ultraviolet divergences of the integral. 
Unlike the local free-energy and condensate, 
the current is an odd function of the
$x$-coordinate, 
and this can be made manifest by breaking the 
$\D$-integral into positive and negative regions, 
and invoking the change of variables 
$\D \to - \D$
in the latter contribution. 
Carrying out this procedure results in 
\begin{equation}
J_y^\p(x)
=
- 2 e \, \l \, \c C_d \, \tfrac{d{-}2}{2}
\int_\m^\infty dz_0 \, (z_0^2 {-} \m^2)^{\frac{d{-}3}{2}} \int_0^1 d\D \, (1{-}\D^2)^{\frac{d{-}4}{2}}
\, \c J \big(z_0, \d(\D) \big| x \big) 
,\end{equation}
where the pre-factor 
$\c C_d$
is that given in 
Eq.~\eqref{eq:Cdpre} 
and the auxiliary function 
$\c J$
is defined to be
\begin{multline}
\c J (z_0, \d | x )
=
\frac{\G(z{+}\frac{1}{2}{+}\g) \G(z{+}\frac{1}{2}{-}\g)}
{2 z \, \G(z{+}b
\d)\G(z{-}b
\d)}
\\
\times
\left[
\Big( z \d
{+} b (1{-}2\x) \Big)
\Psi_1(x) \Psi_3(x)
{-} 
\Big( 
z \d
{-} b (1{-}2\x) \Big)
\Psi_1({-}x) \Psi_3({-}x)
\right]
.\end{multline}
The logarithmic divergence of the integral must be canceled by the counter-term. 
To this end, 
we write
Eq.~\eqref{eq:h}
in the form
\begin{equation}
h
=
-
\l^4 \, \c C_d
\, \tfrac{d-2}{2}
\int_\m^{\infty} \frac{dz_0}{
2 \,
z_0^2} 
(z_0^2 {-} \m^2)^{\frac{d-3}{2}} \int_0^1 d\D \, (1{-}\D^2)^{\frac{d-4}{2}} \D^2
+
\frac{2}{9(4\p)^2}
+
\frac{\ol h}{3(4\p)^2}
\label{eq:newh}
,\end{equation}
which holds near 
$d{=}4$
dimensions. 
This particular form is chosen by identifying the appropriate term in the 
$z_0 \gg 1$
asymptotic expansion of the integrand. 
The vacuum current can then be written as
\begin{equation}
J_y(x)
=
- \left[1{+} \frac{e^2 \, \ol h}{3 (4 \p)^2} \right]
\frac{dB}{dx} 
+ 
J_y^{\p,r}(x)
,\end{equation}
where the pion contribution to the current now appears in the form
\begin{multline}
J_y^{\p,r}(x)
=
- \x (1{-}\x)(1{-}2\x)
\frac{16 e b / \l^3 }{9 (4\p)^2} 
\\
- 2 e \, \l \, \c C_d \, \tfrac{d{-}2}{2}
\int_\m^\infty dz_0 \, (z_0^2 {-} \m^2)^{\frac{d{-}3}{2}} \int_0^1 d\D \, (1{-}\D^2)^{\frac{d{-}4}{2}}
\, 
\overline{\c J} \big(z_0, \d(\D) \big| x \big)
,\end{multline}
with the subtracted function defined by 
\begin{equation}
\overline{\c J}
\big(z_0, \d(\D) \big| x \big)
=
 \c J \big(z_0, \d(\D) \big| x \big) - \frac{2 b \x (1{-}\x)(1{-}2\x) \D^2}{z_0^2}
.\end{equation}

While the logarithmic divergence has been removed by this subtraction, 
there are still power-law divergences that are being regulated in 
$d$ dimensions. 
This behavior must also be added and subtracted in the regulated integral, 
with the added contribution evaluated in dimensional regularization with subsequent analytic continuation required to enable the 
$\e \to 0$
limit to be performed. 
Problematic terms in the 
$z_0 \gg 1$ 
asymptotic expansion of 
$\overline{\c J}$
are
\begin{equation}
j(z_0, \D | x)
=
b (1{-}2\x)
\left\{
1{-}\D^2
+
\frac{\m^2 \D^2 +2 b^2 \x (1{-}\x)
\big[ 1 {-} 6 \D^2 + 5 \D^4 \big]}
{z_0^2}
\right\}
,\end{equation}
where higher-order terms are proportional to 
$z_0^{-4}$, 
which are finite in the ultraviolet. 
In dimensional regularization, 
the integral evaluates to
\begin{eqnarray}
j(x)
&\equiv& 
\int_\m^\infty dz_0 
\, (z_0^2 {-} \m^2)^{\frac{d-3}{2}} 
\int_0^1 d\D \, (1{-}\D^2)^{\frac{d-4}{2}}
\, j(z_0, \D | x)
\notag \\
&=&
- \frac{4}{15} b^3 \x (1{-}\x) (1{-}2\x)
.\end{eqnarray}
The renormalized vacuum current is thus given by the expression
\begin{equation}
J_y^{\p,r}(x)
=
- \frac{8 e / \l^{3}}{(4\p)^2}
\Bigg\{
\overline{j \,}(x)
+
\int_\m^\infty dz_0 \, (z_0^2 {-} \m^2)^{\frac{1}{2}} \int_0^1 d\D 
\Big[
\overline{\c J} \big(z_0, \d(\D) \big| x \big) 
-
j(z_0, \D | x )
\Big]
\Bigg\}
\label{eq:J2}
,\end{equation}
where
\begin{equation}
\overline{j\,}(x)
=
j(x)
+ 
\x (1{-}\x)(1{-}2\x)
\frac{2 b}{9} 
,\end{equation}
includes the integral of the subtraction function and the additional finite contribution arising from 
Eq.~\eqref{eq:newh}.

\begin{figure}[tbp]
\centering 
\includegraphics[width=.495\textwidth]{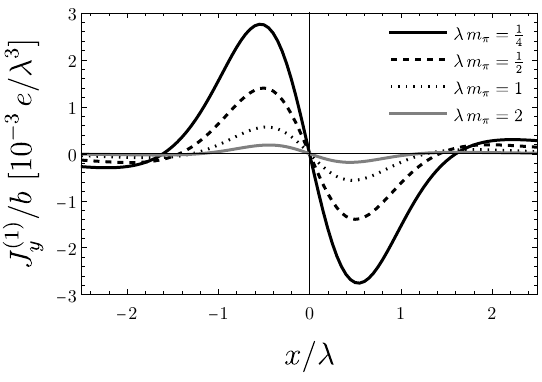}
\caption{\label{f:linear}%
Linear response of the renormalized vacuum current to the strength of the magnetic field. 
The first derivative of the vacuum current with respect to the field strength
$\frac{d}{db} J_y^{\p,r} \big|_{b = 0} = J_y^{(1)} / \, b$
is plotted as a function of 
$x$, 
for a few values of the width
$\l$. 
Unlike the corresponding Maxwell current, 
this renormalized contribution arises solely from charged-pion vacuum fluctuations. 
The linear response is determined by the vacuum polarization tensor and therefore provides a probe of the 
QCD vacuum polarization in an inhomogeneous magnetic field.}
\end{figure}

The renormalized vacuum current has contributions of linear order in the strength of the magnetic field. 
These contributions must be subtracted in order to compare with the locally constant approximation to the 
current. 
The linear contribution to the current arises as a projection of the underlying vacuum polarization response function%
~\cite{Gies:2011he}
\begin{equation}
J_y^{(1)}(x)
=
- 
\int dx' \,
\Pi_{yy}(x,x') A_y(x')
,\end{equation}
which is denoted by 
$\Pi_{\mu\nu}(x,x')$. 
As this kernel is nonlocal, 
the coordinate dependence of the linear contribution to the current need not be proportional to 
$dB/dx$. 
The situation is analogous to the quadratic response in the free-energy density detailed in Sec.~\ref{sec:freee}. 
Unlike the free-energy density, 
there is no ambiguity in defining the vacuum current. 
Additionally, 
the separation of the linear term is required to make contact with the locally constant approximation. 
To this end, 
we define the nonlinear induced vacuum current
\begin{equation}
\c J_y^{\p,r}(x)
= 
J_y^{\p,r}(x) 
- 
J_y^{(1)}(x)
,\end{equation}
by subtracting the contributions that are linear in the magnetic field strength.

The linear response of the vacuum current to the magnetic field strength
$J_y^{(1)}(x)$ 
is essentially the 
current induced from the QCD vacuum polarization. 
The behavior of this linear response is shown in 
Fig.~\ref{f:linear}. 
The response generally diminishes with increasing magnetic-field width, 
reflecting the fact that the renormalized vacuum polarization is induced by spatial variation in the magnetic field and therefore vanishes in the uniform-field limit.

\begin{figure}
\centering 
\includegraphics[width=.495\textwidth]{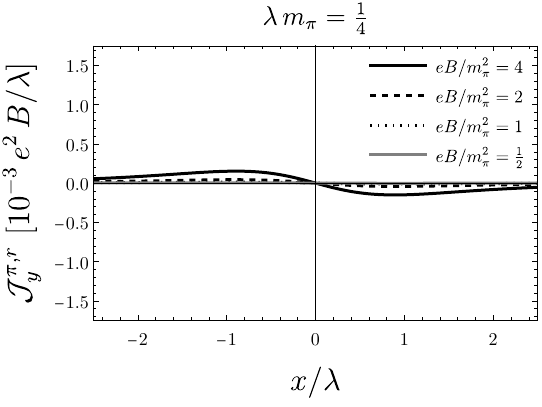}
\hfill
\includegraphics[width=.495\textwidth]{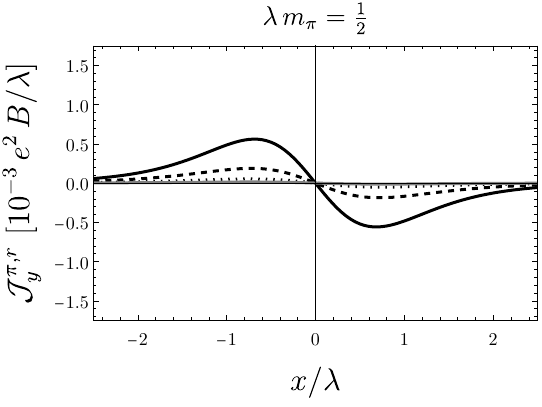}
\\
\includegraphics[width=.495\textwidth]{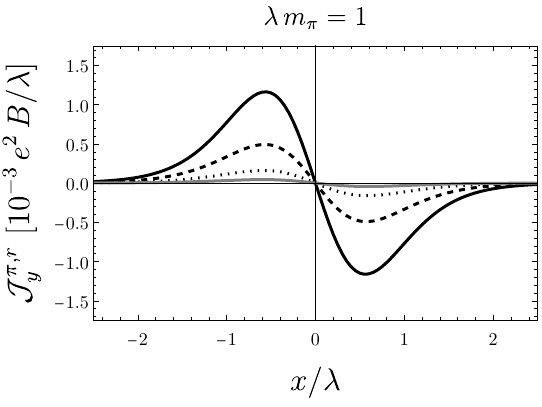}
\hfill
\includegraphics[width=.495\textwidth]{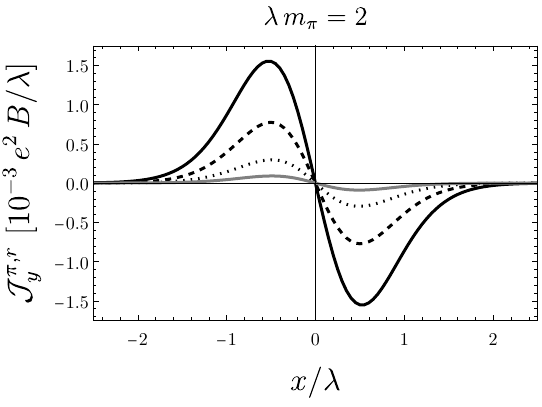}
\caption{\label{f:localmagscaled}%
Spatial variation of the renormalized vacuum current 
$\c J_y^{\p,r}$
in units of 
$e^2 B / \l$. 
The field-strength legend applies to all four plots.
With the linear response subtracted,  
this current encodes the nonlinear response of the QCD vacuum to the field inhomogeneity. 
Due to the small size of the non-linear response, 
the current is remarkably linear for small-width magnetic fields. 
Departure from the linear-response approximation becomes progressively more visible with increasing magnetic-field strength.}
\end{figure}

With the linear response removed, 
the nonlinear induced vacuum current 
$\c J^{\p,r}_y(x)$
is shown in 
Fig.~\ref{f:localmagscaled}.
As such,
it encodes the higher-order electromagnetic response functions of the QCD vacuum to the inhomogeneity. 
As the effect vanishes in a uniform magnetic field, 
we cannot assess the relative size of the current and have directly plotted its value, 
but in units other than the natural scale
$e / \l^3$
that appears in 
Eq.~\eqref{eq:J2}. 
In the chiral limit, 
the vacuum current scales as 
$J_y^{\p, r} \propto e^2 B / \l$, 
and this is the scale we choose in the plot. 
For widths that are small compared to the pion Compton wavelength 
$\l \, m_\p < 1$, 
the current is linearly proportional to the overall strength of the magnetic field to a good approximation,
but this linear contribution has been removed from 
$\c J^{\p,r}_y$. 
The nonlinear response in the field strength is more visible for 
$\l \, m_\p \sim 1$, 
as demonstrated in 
Fig.~\ref{f:localmagscaled}; 
whereas, 
for small-width fields, 
the nonlinear response is only discernible in the largest magnetic field strengths shown.

\begin{figure}
\centering 
\includegraphics[width=.495\textwidth]{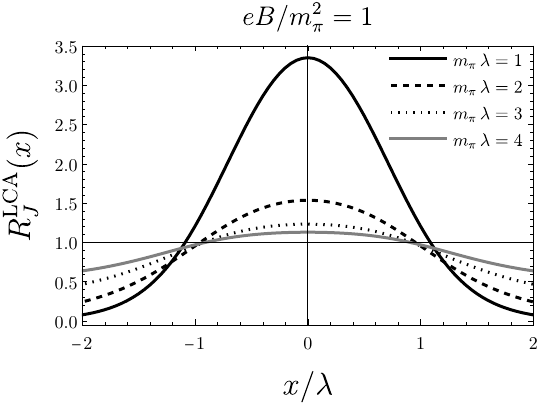}
\hfill
\includegraphics[width=.495\textwidth]{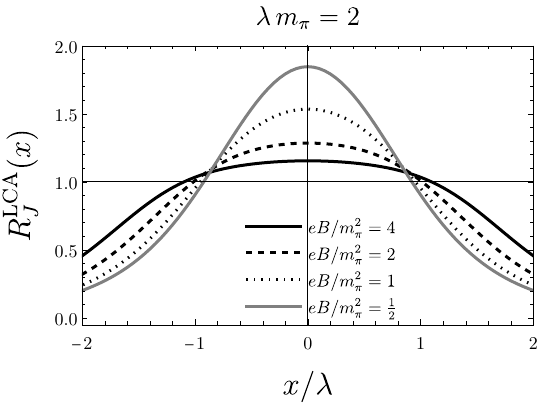}
\\
\caption{\label{f:LCAmag}%
Locally constant approximation to the vacuum current. 
On the left, 
the ratio 
$R^\text{LCA}_{J}(x)$
in 
Eq.~\eqref{eq:LCAmag}
is plotted as a function of 
$x$
using differing widths
$\l$
of the magnetic field, 
for a fixed magnetic field strength
$eB$.
The locally constant approximation describes the induced vacuum current when the ratio is near unity, 
and the approximation becomes exact in the large-width limit.  
On the right, 
the same ratio is plotted for differing magnetic field strengths that share the same spatial width.  
The locally constant approximation also improves with increasing field strength. 
} 
\end{figure}

For large-width magnetic fields, 
the nonlinear vacuum current should be described by a locally constant approximation. 
Using the renormalized magnetization in a uniform magnetic field 
Eq.~\eqref{eq:bulkM}, 
the locally constant approximation is obtained by replacing 
$B \to B(x)$
and taking the curl of the resulting magnetization
\begin{equation}
J_y^\text{LCA}(x)
=
\frac{d}{dx} {\c M}^r \Big( B(x) \Big)
.\end{equation}
The ratio of the nonlinear current with its locally constant approximation 
\begin{equation}
R_J^\text{LCA}(x)
=
\frac{J_y^\text{LCA}(x)}{\c J_y^{\p,r}(x)}
\label{eq:LCAmag}
,\end{equation}
is shown in 
Fig.~\ref{f:LCAmag}. 
The ratio becomes unity when the width of the field is suitably large. 
The pattern shown in the figure mirrors what is found for the locally constant approximation to the chiral 
condensate. 
For the current, 
the locally constant approximation overestimates the size of the current where the magnetic field 
is largest, 
and underestimates the current where the magnetic field is the smallest. 
Due to the crossover in this behavior, 
the approximation does very well in the region where 
$|x| \sim \l$. 
The vacuum current has its extrema for 
$|x| \sim \frac{1}{2} \l$. 
In absolute terms, 
the locally constant approximation consequently does a good job at characterizing the vacuum current where 
its value is appreciable. 
In a fixed magnetic field strength, 
the figure shows marked improvement in the approximation with increasing width. 
Additionally,
the approximation improves with increasing field strength, 
and this trend is also depicted in the figure.

\section{Conclusion}
\label{sec:conc}

The response of the QCD vacuum to an inhomogeneous magnetic field 
is studied using chiral perturbation theory. 
We provide a careful treatment of the effective action using dimensional regularization to arrive at 
ultraviolet-finite quantities, 
for which a single counter-term is required for the renormalization. 
Specifically investigated are the renormalized free-energy density, 
the modification of the chiral condensate, and the induced vacuum current. 
These quantities are explored as a function of the width 
$\l$
of the magnetic field, 
both as integrated responses, 
as well as local responses. 
Through the latter quantities, 
pion fluctuations are demonstrated to respond non-locally to the spatial profile of the magnetic field. 
Only in the limit of a large-width magnetic field is a locally constant approximation to the external 
field valid. 
In this limit, 
the field varies slowly on the scale of the pion Compton wavelength
$\l \gg m_\p^{-1}$, 
and pions respond to the local value of the magnetic field. 
The principal results of this work are obtained for
$\l \sim m_\p^{-1}$. 
In this regime, 
pion fluctuations occur over the same length scale as the magnetic field, 
and 
consequently the response of the vacuum is intrinsically nonlocal. 
The most striking manifestation of such non-locality is the induced vacuum current, 
which identically vanishes in a uniform magnetic field and therefore serves as a probe of
the field inhomogeneity.

The quantities obtained in this work are determined non-perturbatively in both the width 
$\l  \sim m_\p^{-1}$
and strength of the magnetic field 
$e B \sim  m_\p^2$. 
The weak-field regime, 
however, 
naturally motivates the determination of the underlying response functions of the QCD vacuum.
In this investigation, 
we find a finite linear vacuum current after renormalization of the Maxwell source current, 
see Fig.~\ref{f:linear}. 
This current is induced by the QCD vacuum polarization, 
through a field-profile projection of the nonlocal polarization tensor.
A more general quantity is the vacuum polarization in an external field
\begin{equation}
\Pi_{\m \n}(x,x')
= 
\frac{\d^2 \, \G[\c A]}{\d \c A_\m(x) \d \c A_\n(x')} 
\Bigg|_{\c A = A}
.\end{equation}
While the present work focuses on specific observables derived from the effective action, 
the techniques developed here can be extended toward determining the 
full vacuum polarization tensor in an inhomogeneous magnetic background. 
Such a calculation would provide a direct characterization of the nonlocal electromagnetic 
response of the QCD vacuum.
Additionally, 
the free-energy density contains a finite quadratic response to the magnetic field 
after renormalization of the Maxwell energy density, 
see Figs.~\ref{f:susc} and \ref{f:quadratic}. 
This effect owes to an underlying nonlocal magnetic susceptibility, 
which is defined by 
\begin{equation}
\chi_B(x,x')
= 
\frac{\d^2 \, \G [A]}{\d B(x) \d B(x')}
.\end{equation}
Although difficult to compute, 
the susceptibility kernel contains information beyond that encoded in integrated observables, 
and would provide a direct characterization of the spatial extent of the QCD vacuum response 
to magnetic fields.
The determination of these nonlocal response functions remains an important direction for future work, 
and may provide a more complete understanding of vacuum structure in inhomogeneous backgrounds.
The analytical framework developed here also suggests a path toward the study of a class of solvable
time-dependent external fields. 
Extending these methods to spacetime-dependent backgrounds would enable investigations of dynamical vacuum polarization and the real-time response of the QCD vacuum to electromagnetic fields.

\appendix

\section{Integrals of the Product of Continuum Eigenfunctions}
\label{s:integrals}

To compute quantities that are integrated responses, 
the trace of the Green's function is required. 
In the quantum mechanical problem, 
this translates into a spatial integral of the product of two 
continuum eigenfunctions. 
A few methods applicable to such integrals are described in 
Ref.~\cite{Buchholz}. 
Here, 
we provide a sketch of the derivation required above for the inhomogeneous magnetic field.

Two continuum eigenfunctions 
$\Psi_a(x|\c E_a)$
and
$\Psi_b(x|\c E_b)$
are taken to satisfy the  
Schr\"odinger equation 
\begin{equation}
\left[
- \frac{d^2}{dx^2} + V(x) 
\right] 
\Psi(x|\c E) 
= 
\c E \,
\Psi(x| \c E)
,\end{equation}
where the energies lie in the continuous spectrum. 
Note that 
$\c E = - E$, 
where
$E = p_3^2 + p_4^2 + \cdots$
is employed in the main text. 
For a uniform magnetic field, 
the potential depends on 
$p_2$
and 
$e B$; 
whereas, 
the inhomogeneous magnetic field additionally introduces 
$\l$. 
For solutions sharing these parameters,
but with different energies
$\c E_a$
and
$\c E_b \equiv \c E$, 
we have the indefinite integral
\begin{equation}
\big(
\c E_a - \c E 
\big)
\int
\Psi_a(x|\c E_a) \Psi_b (x| \c E)
\, dx
=
\Psi_a(x| \c E_a)  
\Psi'_b (x|\c E)
-
\Psi'_a(x|\c E_a) 
\Psi_b (x| \c E)
+
\c C
,\end{equation}
where the right-hand side results from application of the Schr\"odinger equation, 
and primes denote differentiation with respect to 
$x$.  
While independent of 
$x$, 
the form of the integration constant 
$\c C$
is not completely arbitrary with respect to energy. 
Given the behavior of the left-hand side, 
the right-hand side must vanish at least linearly in the limit 
$\c E_a \to \c E$. 
Thus, we must have the $x$-independent integration constant of the form
\begin{equation}
\c C = W[\Psi_a,\Psi_b]\Big|_{\c E} + (\c E_a - \c E) \, C
,\end{equation} 
where
$W$
denotes the Wronskian of the solutions,
and 
$C$
is a non-singular function of 
$\D \c E = \c E_a - \c E$
at 
$\D \c E = 0$. 
Expanding both sides to first order in 
$\D \c E$
leads to the indefinite integral
\begin{equation}
\int
\Psi_a(x|\c E) \Psi_b (x|\c E)
\, dx
=
\frac{\partial \Psi_a(x|E)}{\partial \c E}  
\Psi'_b (x| \c E)
-
\frac{\partial \Psi'_a(x|\c E)}{\partial \c E} 
\Psi_b (x| \c E)
+
C
\label{eq:indefinite}
,\end{equation}
where 
$C$
is now technically evaluated at 
$\D \c E = 0$.

The trace of the one-dimensional Green's function requires integrating over all 
$x$; 
however,
the resulting integral does not converge because the regular solutions at 
$x = \pm \infty$
are irregular at 
$x = \mp \infty$.
Defining the otherwise divergent integral with an infrared cutoff 
$L$
\begin{equation}
\c I
\equiv 
\c I(L)
=
\int_{-\frac{L}{2}}^{+\frac{L}{2}}
\Psi_a(x|\c E) \Psi_b (x| \c E)
\, dx
,\end{equation}
we have the result for the definite integral from 
Eq.~\eqref{eq:indefinite}
\begin{multline}
\c I
=
\frac{\partial \Psi_a\big(\frac{L}{2}\big|\c E\big)}{\partial \c E}  
\Psi'_b \big(\tfrac{L}{2} \big| \c E \big)
-
\frac{\partial \Psi'_a \big(\frac{L}{2}\big|\c E\big)}{\partial \c E}  
\Psi_b \big(\tfrac{L}{2}\big| \c E\big)
\\
-
\frac{\partial \Psi_a \big({-}\frac{L}{2}\big|\c E\big)}{\partial \c E}  
\Psi'_b \big({-}\tfrac{L}{2}\big| \c E\big)
+
\frac{\partial \Psi'_a \big({-}\frac{L}{2} \big| \c E \big)}{\partial \c E}  
\Psi_b \big({-}\tfrac{L}{2} \big| \c E \big)
\label{eq:definite}
.\end{multline}

To apply 
Eq.~\eqref{eq:definite} 
to the inhomogeneous magnetic field problem, 
we take the two solutions to be
$\Psi_1(x)$
and
$\Psi_3(x)$
appearing in 
Eqs.~\eqref{eq:psi1} and \eqref{eq:psi3}. 
Note that 
$x = \frac{L}{2}$
becoming large corresponds to 
$\x = 1 - \epsilon$
for 
$\epsilon \ll 1$, 
whereas 
$x = - \frac{L}{2}$
for 
$L \to \infty$
corresponds to 
$\x  = \epsilon$
for 
$\epsilon \ll 1$. 
The conversion is given by 
$\epsilon = 2 e^{- L}$.
Using the endpoint behavior of the hypergeometric wavefunctions and their derivatives, 
one arrives at the result
\begin{equation}
\frac{\c I}{W}
=
- 
\frac{\l^2}{8}
\left( \frac{1}{\a} {+} \frac{1}{\be} \right)
\left[ 
\psi \left( z {+} \tfrac{1}{2} {+} \g \right) + \psi \left( z {+} \tfrac{1}{2} {-} \g \right) 
\right]
+ 
f_\epsilon(\a) + f_\epsilon(\be)
\label{eq:integral}
,\end{equation}
where 
$W$
is the Wronskian in 
Eq.~\eqref{eq:W}, 
and 
$\psi(z)$
is 
Euler's digamma function. 
The infrared divergent contributions are contained in the auxiliary function 
$f_\epsilon(\chi)$, 
which has the form 
\begin{equation}
f_\epsilon(\chi)
=
- \frac{\l^2}{8\chi}
\big[ 
\log \epsilon
-
\psi(2\chi{+}1)
-
\psi(2\chi)
\big]
.\end{equation}
The result in 
Eq.~\eqref{eq:integral}
is a factor of two smaller than that quoted in 
Ref.~\cite{Cangemi:1995ee}, 
and the function 
$f_\epsilon(\chi)$
is not reported in any regularization scheme. 
The logarithmic divergent terms appearing in 
Eq.~\eqref{eq:integral}, 
however,
exactly match those found from analyzing the endpoint behavior of the integral in 
Eq.~\eqref{eq:Trace1D}, 
which corroborates the factor of two. 
As the analogous result for the effective action computed in 
$d{=}3$
dimensions agrees with that obtained in
Ref.~\cite{Cangemi:1995ee}, 
the factor of two for their integral must be a typo.

\section{Integration of the Local Free-Energy Density}
\label{s:integrate}

An important check of all results is the verification that local quantities indeed produce the integrated response when 
integrated over all space. 
As the chiral condensate can be obtained from the free-energy, 
it suffices to consider the local free-energy density. 
Here, 
we verify that the integrated response
$\overline{\c F}$
in
Eq.~\eqref{eq:FF}
agrees with the spatial integral of the local free-energy density
Eq.~\eqref{eq:fBlocal}. 
This proves to be a non-trivial task.

In obtaining the integrated free-energy density, 
a shift of the $p_2$ integration variable is utilized to arrive at 
Eq.~\eqref{eq:Fz}. 
This shift is carried out on a particular subset of terms of the integrand, 
which is only possible because each term is a finite integral when regulated. 
We can obtain an alternate formula for the integrated free-energy density without shifting 
$p_2$. 
Starting with 
Eq.~\eqref{eq:FBTrace1D}, 
this results in the two-dimensional integral
\begin{multline}
\overline{ \c F}^\p
=
\l \, \c C_d
\int_\m^{\infty} d z_0 \, (z_0^2{-}\m^2)^{\frac{d-1}{2}} \int_0^1 d \D \, (1{-}\D^2)^{\frac{d-2}{2}}
\Bigg\{
\psi\left(z{+}\tfrac{1}{2}{+}\g\right)
+
\psi\left(z{+}\tfrac{1}{2}{-}\g\right)
\\
- \frac{1}{z}
\left[  \frac{z^2 {+} b^2 
\d^2}{z^2 {-} b^2 
\d^2} + (z {-} b 
\d) \psi(z{+} b
\d) + (z{+} b
\d) \psi(z{-} b
\d)  \right]
\Bigg\}
\label{eq:bulkalt}
,\end{multline}
where we have employed the transformation from 
$p_2$ 
and
$E$
to the variables
$z_0$
and
$\D$, 
as in 
Section~\ref{sec:freee}, 
and the multiplicative factor
$\c C_d$
is that given in 
Eq.~\eqref{eq:Cdpre}.

The logarithmic divergence of 
Eq.~\eqref{eq:bulkalt}
is canceled by the addition of the counter-term, 
written in the form of  
Eq.~\eqref{eq:hlocal}.
Even after this term is added, 
the bulk free-energy density above still possesses a quadratic divergence that is regulated in 
$d$ dimensions. 
As in Section~\ref{sec:freee}, 
we add and subtract the  
$z_0 \gg 1$
asymptotic behavior of the integrand. 
The subtraction cures the power-law divergence, 
whereas the added term is rendered finite by analytic continuation to 
$d{=}4$
dimensions. 
Carrying out this procedure on 
Eq.~\eqref{eq:bulkalt}, 
leads to a result that can be expressed in the form
\begin{equation}
\overline{\c F}
= 
\frac{4 \l}{3}
\left[
1+ \frac{e^2 \, \ol h}{3 (4 \p)^2}
\right]
\frac{B^2}{2} 
+ 
\int_{-\infty}^{+\infty} \c F_B(x) \, dx 
\label{eq:Falt}
,\end{equation}
where the spatial integral is that of the (partially) 
renormalized local magnetic free-energy density 
$\c F_B(x)$
precisely as obtained in 
Eq.~\eqref{eq:fBlocal}. 
On account of this formula,  
integration of the local free-energy density does produce the integrated free-energy density, 
however, 
in terms of a two-dimensional integral, 
rather than the one-dimensional integral appearing in 
Eq.~\eqref{eq:Frenorm}. 
The relative simplicity of that result necessitates the shift of the 
$p_2$
variable that enables all but one integral to be performed analytically. 
We do not believe, 
however,  
that it is possible to perform the analogous integration for local quantities, 
such as in
Eq.~\eqref{eq:fintermed}, 
which have an intricate dependence on 
$p_2$
(or equivalently on the variable
$\D$). 
Consequently, 
we are unable to convert the two-dimensional integral analytically into a one-dimensional integral 
after renormalization.

The equality of these one-dimensional and two-dimensional integrals for the 
integrated free-energy density appears highly non-trivial. 
The difference of 
Eqs.~\eqref{eq:Falt} 
and
\eqref{eq:FF}
can be expressed as
\begin{equation}
\D \overline{\c F} 
= 
\frac{4}{(4\p)^2 \l^3}
\, \D \overline{f \,}
,\end{equation}
where, 
removing the overall dimensionful pre-factor, 
we have
\begin{multline}
\D \overline{f \,}
=
\frac{b^2}{45} - \frac{2 \, b^4}{35}
+
\int_\m^\infty dz_0 (z_0^2 {-} \m^2)^{\frac{3}{2}} \int_0^1 d\D (1{-}\D^2)
\Bigg\{
\psi(z_0{+}1) + \psi(z_0) 
+ 
\ol g(z_0, \D)
\\
-\frac{1}{z}
\left[  \frac{z^2 {+} b^2 
\d^2}{z^2 {-} b^2 
\d^2} + (z {-} b 
\d) \psi(z{+} b
\d) + (z{+} b
\d) \psi(z{-} b
\d)  \right]
\Bigg\}
\label{eq:DFbar}
.\end{multline}
The subtraction term 
$\ol g(z_0,\D)$
appearing in the integrand is the dimensionless spatial integral of 
Eq.~\eqref{eq:littleg}, 
namely
\begin{eqnarray}
\ol g(z_0,\D) 
&\equiv&
\frac{1}{\l} \int_{-\infty}^{+\infty} g(z_0, \D | x) \, dx
\notag \\
&=&
\frac{b^2}{z_0^2} (1{-}3\D^2)
{+} 
\frac{b^2}{6z_0^4}
\left[
1{-}5\D^2 {+} 18 \m^2 \D^2 {-} b^2 (3 {-}18 \D^2 {+} 7 \D^4)
\right]
.\end{eqnarray}
We have verified 
$\D \overline{f \,} = 0$  
numerically, 
for a range of parameters 
$b$ 
and 
$\m$
on the order of unity.
As an additional check, 
we investigate the series expansion of
$\D \overline{f \,}$
in powers of 
$b^2$, 
which enables analytic computation of the integral over 
$\D$
order-by-order in the expansion. 
At second order, 
we have
\begin{multline}
\D \overline{f \,}^{(2)}
=
\frac{b^2}{15}
\Bigg\{
\frac{1}{3}
+
\int_{\m}^\infty dz_0 (z_0^2 {-} \m^2)^{\frac{3}{2}}
\Bigg[
\frac{4 z_0^3 {+} z_0^2 {+} 6 z_0 \m^2 {+} 4 \m^2}{z_0^5}
\\
-
2\frac{3 z_0^2{+}2 \m^2}{z_0^3} \psi^{(1)}(z_0)
-
2\frac{z_0^2{-}\m^2}{z_0^2} \psi^{(2)}(z_0)
\Bigg]
\Bigg\}
,\end{multline}
where
$\psi^{(n)}(z) = \frac{d^n}{dz^n} \psi(z)$
is used to denote the poly-gamma function. 
The integral appearing above numerically evaluates to 
$- \frac{1}{3}$
to high precision for values of 
$\m$ 
on the order of unity, 
leading to 
$\D \overline{f \, }^{(2)} = 0$. 
At fourth order in the expansion of 
Eq.~\eqref{eq:DFbar}, 
we have
\begin{multline}
\D \overline{f \,}^{(4)}
=
\frac{b^4}{420}
\Bigg\{
-24
+
\int_\m^\infty dz_0 (z_0^2 {-}\m^2)^{\frac{3}{2}}
\Bigg[
\frac{15 z_0^4 {-} 72 z_0^2 \m^2 {-} 48 \m^4}{z_0^9}
+
14 \frac{z_0^2{+}4 \m^2}{z_0^5} \psi^{(1)}(z_0)
\\
-
14 \frac{z_0^2 {+} 4 \m^2}{z_0^4} \psi^{(2)}(z_0)
-
4 \frac{(z_0^2{-}\m^2)(5z_0^2 {+} 2 \m^2)}{z_0^5} \psi^{(3)}(z_0)
-
2 \frac{(z_0^2{-}\m^2)^2}{z_0^4}\psi^{(4)}(z_0)
\Bigg]
\Bigg\}
.\end{multline}
The integral numerically evaluates to 
$24$ 
to high precision using values of 
$\m$ 
on the order of unity, 
leading to 
$\D \overline{f \,}^{(4)} = 0$. 
As higher-order terms become increasingly cumbersome to display,
we present only the sixth-order term
\begin{multline}
\D \overline{f \,}^{(6)}
=
- \frac{b^6}{22680}
\int_\m^\infty dz_0 (z_0^2{-}\m^2)^{\frac{3}{2}}
\Bigg[
45 \frac{35 z_0^6 {-} 60 z_0^4 \m^2 {-} 48 z_0^2 \m^4 {-} 32 \m^6}{z_0^{13}}
- 378 \frac{z_0^2 {-} 6 \m^2}{z_0^7} \psi^{(1)}(z_0)
\\
+ 378 \frac{z_0^2 {-} 6 \m^2}{z_0^6} \psi^{(2)}(z_0)
- 18 \frac{16 z_0^4 {-} 39 z_0^2 \m^2 {-} 12 \m^4}{z_0^7} \psi^{(3)}(z_0)
+ 54 \frac{3 z_0^4 {+} z_0^2 \m^2 {-}4 \m^4 }{z_0^6} \psi^{(4)}(z_0)
\\
+ 6 \frac{(z_0^2{-}\m^2)^2 (7 z_0^2 {+} 2 \m^2)}{z_0^7} \psi^{(5)}(z_0)
+ 2 \frac{(z_0^2{-}\m^2)^3}{z_0^6} \psi^{(6)}(z_0)
\Bigg]
.\end{multline} 
This integral numerically evaluates to zero for 
$\m$
on the order of unity.%
\footnote{
In practice, 
a numerically stable evaluation requires a cutoff on 
$z_0$, 
above which the integral is performed analytically. 
The same method is employed for the numerical evaluation of
Eq.~\eqref{eq:DFbar}, 
as well as all other integrals that are evaluated in this work. 
Cutoff independence serves as a crucial check on the stability of numerical results.
We have also verified that the same numerical results are obtained for very large values of the cutoff
by using variables of octuple precision. 
}
While we have been unable to discern the subtle pattern giving rise to the vanishing of 
Eq.~\eqref{eq:DFbar}, 
its vanishing is required due to the fact that
Eqs.~\eqref{eq:Falt} and \eqref{eq:FF}
are different expressions for the same quantity.

\acknowledgments

P.A.~is supported by the 
U.S.~Department of Energy, 
Office of Science, 
Office of Nuclear Physics and Quantum Horizons Program under Award Number 
DE-SC$0024385$. 
P.A. acknowledges the hospitality of the Kavli Institute for Theoretical Physics, Santa Barbara, 
through which the research was supported in part by the U.S. National Science Foundation under 
Grant No.~NSF PHY-$1748958$.
B.C.T.~acknowledges the hospitality of the IAS-$4$ group at the Forschungszentrum J\"ulich, 
through which the research was supported in part by funds provided by the Excellence Network Initiative at the Forschungszentrum J\"ulich under FK EXNET-$01$-$07$.

\bibliographystyle{JHEP}
\bibliography{bibly}

\end{document}